\documentclass[journal,draftclsnofoot,onecolumn]{IEEEtran}
\usepackage{graphicx}
\usepackage{amssymb}
\usepackage{color}

\usepackage{caption}
\renewcommand{\theequation}{\arabic{section}.\arabic{equation}}
\newtheorem{theorem}{Theorem}

\newtheorem{definition}{Definition}

\newtheorem{remark}{Remark}

\begin{document}

\title{Multiple-Access Relay Wiretap Channel}

\author{Bin~Dai
        and~Zheng~Ma

\thanks{B. Dai and Z. Ma are with the
School of Information Science and Technology,
Southwest JiaoTong University, Chengdu 610031, China
e-mail: daibin@home.swjtu.edu.cn, zma@home.swjtu.edu.cn.}
}

\maketitle

\begin{abstract}

In this paper, we investigate the effects of an
additional trusted relay node on the secrecy of multiple-access wiretap
channel (MAC-WT) by considering the model of multiple-access
relay wiretap channel (MARC-WT). More specifically,
first, we investigate the discrete memoryless MARC-WT. Three
inner bounds (with respect to decode-forward (DF), noise-forward (NF) and compress-forward
(CF) strategies) on the secrecy capacity region are
provided. Second, we investigate the degraded discrete memoryless MARC-WT, and present an outer bound on the secrecy capacity region of this degraded model.
Finally, we investigate the Gaussian MARC-WT, and
find that the NF and CF strategies help to enhance Tekin-Yener's achievable
secrecy rate region of Gaussian MAC-WT. Moreover, we find that if the channel from the
transmitters to the relay is less noisy than the channels
from the transmitters to the legitimate receiver and the wiretapper, the achievable
secrecy rate region of the DF strategy is even larger than the corresponding regions of the NF and CF strategies.

\end{abstract}


\section{Introduction \label{secI}}

Equivocation was first introduced into channel coding by
Wyner in his study of wiretap channel \cite{Wy}. It is a kind of discrete
memoryless degraded broadcast channels. The objective is
to transmit messages to the legitimate receiver, while keeping
the wiretapper as ignorant of the messages as possible. Based
on Wyner¡¯s work, Leung-Yan-Cheong and Hellman studied
the Gaussian wiretap channel (GWC) \cite{CH}, and showed that its
secrecy capacity was the difference between the main channel
capacity and the overall wiretap channel capacity (the cascade
of main channel and wiretap channel).

After the publication of Wyner's work, Csisz$\acute{a}$r and K\"{o}rner \cite{CK} investigated a more general situation: the broadcast
channels with confidential messages (BCC). In this model, a common message and a confidential message were
sent through a general broadcast channel. The common message was assumed to be decoded correctly by the
legitimate receiver and the wiretapper, while the confidential message was only allowed to be obtained by the
legitimate receiver. This model is also a generalization of \cite{KM}, where no confidentiality condition is imposed. The
capacity-equivocation region and the secrecy capacity region of BCC \cite{CK} were totally determined, and the results
were also a generalization of those in \cite{Wy}. Furthermore, the capacity-equivocation region of Gaussian BCC was determined in \cite{LPS}.

By using the approach of \cite{Wy} and \cite{CK}, the information-theoretic security for other multi-user communication
systems has been widely studied, see the followings.
\begin{itemize}

\item For the broadcast channel, Liu et al. \cite{LMSY} studied the broadcast channel with two confidential messages (no
common message), and provided an inner bound on the secrecy capacity region. Furthermore, Xu et al. \cite{XCC}
studied the broadcast channel with two confidential messages and one common message, and provided inner
and outer bounds on the capacity-equivocation region.

\item For the multiple-access channel (MAC), the security problems are split into two directions.
\begin{itemize}

\item The first is that two users wish to transmit their corresponding messages to a destination, and meanwhile,
they also receive the channel output. Each user treats the other user as a wiretapper, and wishes to
keep its confidential message as secret as possible from the wiretapper. This model is usually called the
MAC with confidential messages, and it was studied by Liang and Poor \cite{LP}. An inner bound on the
capacity-equivocation region is provided for the model with two confidential messages, and the capacity-equivocation
region is still not known. Furthermore, for the model of MAC with one confidential message
\cite{LP}, both inner and outer bounds on capacity-equivocation region are derived. Moreover, for the degraded
MAC with one confidential message, the capacity-equivocation region is totally determined.

\item The second is that an additional wiretapper has access to the MAC output via a wiretap channel, and
therefore, how to keep the confidential messages of the two users as secret as possible from the additional
wiretapper is the main concern of the system designer. This model is usually called the multiple-access
wiretap channel (MAC-WT). The Gaussian MAC-WT was investigated in \cite{TY2, TY1}. An inner bound on the
capacity-equivocation region is provided for the Gaussian MAC-WT. Other related works on MAC-WT
can be found in \cite{EU, BU, WB, WB1, YA, HKY, XDD}.

\end{itemize}

\item For the interference channel, Liu et al. \cite{LMSY} studied the interference channel with two confidential messages,
and provided inner and outer bounds on the secrecy capacity region. In addition, Liang et al. \cite{LP2} studied
the cognitive interference channel with one common message and one confidential message, and the capacity-equivocation
region was totally determined for this model.

\item For the relay channel, Lai and Gamal \cite{LG} studied the relay-eavesdropper channel, where a source wishes
to send messages to a destination while leveraging the help of a trusted relay node to hide those messages from
the eavesdropper. Three inner bounds (with respect to decode-forward, noise-forward and compress-forward
strategies) and one outer bound on the capacity-equivocation region were provided in \cite{LG}. Furthermore,
Tang et. al. \cite{TLSP} introduced the noise-forward strategy of \cite{LG} into the wireless communication networks,
and found that with the help of an independent interferer, the security of the wireless communication networks is enhanced.
In addition, Oohama
\cite{Oo} studied the relay channel with confidential messages, where a relay helps the transmission of messages from
one sender to one receiver. The relay is considered not only as a sender that helps the message transmission
but also as a wiretapper who can obtain some knowledge about the transmitted messages. Measuring the
uncertainty of the relay by equivocation, the inner and outer bounds on the capacity-equivocation region were
provided in \cite{Oo}.

\end{itemize}

Recently, Ekrem and Ulukus \cite{EU1} investigated the effects of user cooperation on the secrecy of broadcast
channels by considering a cooperative relay broadcast channel. They showed that user cooperation can increase
the achievable secrecy rate region of \cite{LMSY}.

\begin{figure}[htb]
\centering
\includegraphics[scale=0.6]{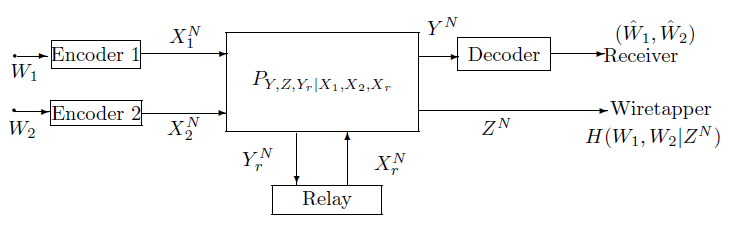}
\caption{The multiple-access relay wiretap channel}
\label{f1}
\end{figure}

In this paper, we study the multiple-access relay wiretap
channel (MARC-WT), see Figure \ref{f1}. This model generalizes
the MAC-WT by considering an additional trusted relay node. The
motivation of this work is to investigate the effects of the trusted relay node on the secrecy of MAC-WT, and whether
the achievable secrecy rate region of \cite{TY1} can be enhanced by
using an additional relay node.

First, we provide three inner bounds on the secrecy capacity
region (achievable secrecy rate regions) of the discrete
memoryless model of Figure \ref{f1}.
The decode-forward (DF), noise-forward (NF) and
compress-forward (CF) relay strategies are used in the construction of the inner bounds.
Second, we investigate the degraded discrete memoryless MARC-WT, and present an outer bound on the secrecy capacity region of this degraded case.
Finally, the Gaussian model of
Figure \ref{f1} is investigated, and we find that with the help of this additional
trusted relay node, Tekin-Yener¡¯s achievable secrecy rate region of the
Gaussian MAC-WT \cite{TY1} is enhanced.

In this paper, random variab1es, sample values and
alphabets are denoted by capital letters, lower case letters and calligraphic letters, respectively.
A similar convention is applied to the random vectors and their sample values.
For example, $U^{N}$ denotes a random $N$-vector $(U_{1},...,U_{N})$,
and $u^{N}=(u_{1},...,u_{N})$ is a specific vector value in $\mathcal{U}^{N}$
that is the $N$th Cartesian power of $\mathcal{U}$.
$U_{i}^{N}$ denotes a random $N-i+1$-vector $(U_{i},...,U_{N})$,
and $u_{i}^{N}=(u_{i},...,u_{N})$ is a specific vector value in $\mathcal{U}_{i}^{N}$.
Let $P_{V}(v)$ denote the probability mass function $Pr\{V=v\}$. Throughout the paper,
the logarithmic function is to the base 2.

The organization of this paper is as follows. Section \ref{secII}
provides the achievable secrecy rate regions of the discrete
memoryless model of Figure \ref{f1}. The Gaussian model of Figure
\ref{f1} is investigated in Section \ref{secIII}. Final conclusions are provided
in Section \ref{secV}.

\section{Discrete memoryless multiple-access relay wiretap channel}\label{secII}
\setcounter{equation}{0}

\subsection{Inner bounds on the secrecy capacity region of the discrete memoryless MARC-WT}\label{sec2.1}

The discrete memoryless model of Figure \ref{f1} is a five-terminal discrete channel consisting of finite sets
$\mathcal{X}_{1}$, $\mathcal{X}_{2}$, $\mathcal{X}_{r}$, $\mathcal{Y}$, $\mathcal{Y}_{r}$, $\mathcal{Z}$ and a transition
probability distribution $P_{Y,Y_{r},Z|X_{1},X_{2},X_{r}}(y,y_{r},z|x_{1},x_{2},x_{r})$. $X_{1}^{N}$, $X_{2}^{N}$ and $X^{N}_{r}$
are the channel inputs from the transmitters and
the relay respectively, while $Y^{N}$, $Y_{r}^{N}$, $Z^{N}$ are the channel outputs at the legitimate receiver, the relay and the wiretapper, respectively.
The channel is discrete memoryless, i.e., the channel outputs $(y_{i},y_{r,i},z_{i})$ at time $i$ only depend on the channel
inputs $(x_{1,i},x_{2,i},x_{r,i})$ at time $i$.

\begin{definition}(\textbf{Channel encoders})\label{def1}
The confidential messages $W_{1}$ and $W_{2}$ take values in $\mathcal{W}_{1}$, $\mathcal{W}_{2}$, respectively.
$W_{1}$ and $W_{2}$ are independent and uniformly distributed over
their ranges. The channel encoders $f_{E1}$ and $f_{E2}$ are stochastic encoders that map the messages $w_{1}$ and $w_{2}$ into the codewords
$x_{1}^{N}\in \mathcal{X}_{1}^{N}$ and $x_{2}^{N}\in \mathcal{X}_{2}^{N}$, respectively.
The transmission rates of the confidential messages $W_{1}$ and $W_{2}$ are
$\frac{\log\|\mathcal{W}_{1}\|}{N}$ and $\frac{\log\|\mathcal{W}_{2}\|}{N}$, respectively.
\end{definition}

\begin{definition}(\textbf{Relay encoder})\label{def2}
The relay encoder $\varphi_{i}$ is also a stochastic encoder that maps the signals $(y_{r,1},y_{r,2},...,y_{r,i-1})$
received before time $i$ to the channel input $x_{r,i}$.

\end{definition}

\begin{definition}(\textbf{Decoder})\label{def3}
The decoder for the legitimate receiver is a mapping $f_{D}: \mathcal{Y}^{N}\rightarrow \mathcal{W}_{1}\times \mathcal{W}_{2}$,
with input $Y^{N}$ and outputs $\hat{W}_{1}$, $\hat{W}_{2}$. Let $P_{e}$ be the error probability of the legitimate receiver,
and it is defined as $Pr\{(W_{1},W_{2})\neq (\hat{W}_{1}, \hat{W}_{2})\}$.

\end{definition}

The equivocation rate at the wiretapper is defined as
\begin{equation}\label{e201}
\Delta=\frac{1}{N}H(W_{1},W_{2}|Z^{N}).
\end{equation}

A rate pair $(R_{1}, R_{2})$ (where $R_{1}, R_{2}\geq 0$) is called
achievable with perfect secrecy if, for any $\epsilon>0$ (where $\epsilon$ is an arbitrary small positive real number), there exists a
sequence of codes $(2^{NR_{1}}, 2^{NR_{2}}, N)$ such that
\begin{eqnarray}\label{e202}
&&\frac{\log\parallel \mathcal{W}_{1}\parallel}{N}=R_{1},
\frac{\log\parallel \mathcal{W}_{2}\parallel}{N}=R_{2}, \nonumber\\
&&\Delta\geq R_{1}+R_{2}-\epsilon, \,\,\,\,P_{e}\leq \epsilon.
\end{eqnarray}
Note that the above secrecy requirement on the full message set also ensures the secrecy of
individual message, i.e., $\frac{1}{N}H(W_{1},W_{2}|Z^{N})\geq R_{1}+R_{2}-\epsilon$ implies that
$\frac{1}{N}H(W_{t}|Z^{N})\geq R_{t}-\epsilon$ for $t=1, 2$, and the proof is as follows.
\begin{IEEEproof}
Since
\begin{eqnarray}\label{e202.x}
&&0\geq R_{1}+R_{2}-\epsilon-\frac{1}{N}H(W_{1},W_{2}|Z^{N})=\frac{1}{N}H(W_{1})+\frac{1}{N}H(W_{2})-\frac{1}{N}H(W_{1},W_{2}|Z^{N})-\epsilon\nonumber\\
&&=\frac{1}{N}H(W_{1})+\frac{1}{N}H(W_{2})-\frac{1}{N}H(W_{1}|Z^{N})-\frac{1}{N}H(W_{2}|W_{1},Z^{N})-\epsilon\nonumber\\
&&\geq \frac{1}{N}H(W_{1})+\frac{1}{N}H(W_{2})-\frac{1}{N}H(W_{1}|Z^{N})-\frac{1}{N}H(W_{2}|Z^{N})-\epsilon\nonumber\\
&&=\frac{1}{N}I(W_{1};Z^{N})+\frac{1}{N}I(W_{2};Z^{N})-\epsilon,
\end{eqnarray}
and $\frac{1}{N}I(W_{1};Z^{N})\geq 0$, $\frac{1}{N}I(W_{2};Z^{N})\geq 0$, it is easy to see that $\frac{1}{N}I(W_{1};Z^{N})\leq \epsilon$, $\frac{1}{N}I(W_{2};Z^{N})\leq \epsilon$,
which implies that $\frac{1}{N}H(W_{t}|Z^{N})\geq R_{t}-\epsilon$ for $t=1, 2$. The proof is completed.
\end{IEEEproof}

The secrecy capacity region $\mathcal{R}^{d}$ is a set composed of
all achievable secrecy rate pairs $(R_{1},R_{2})$. Three inner bounds (with respect to DF, NF and CF strategies)
on $\mathcal{R}^{d}$ are provided in the following Theorem \ref{T1}, \ref{T2}, \ref{T3}.

Our first step is
to characterize the inner bound on the secrecy capacity region $\mathcal{R}^{d}$ by using Cover-El Gamal's Decode and Forward (DF)
Strategy \cite{CG}. In the DF Strategy, the relay node will first decode the confidential messages, and then re-encode them
to cooperate with the transmitters. The superposition coding and random binning techniques
will be combined with the classical DF strategy \cite{CG} to characterize the DF inner bound of the discrete memoryless MARC-WT.
The following Theorem \ref{T1} shows the DF inner bound on $\mathcal{R}^{d}$.

\begin{theorem}\label{T1}
\textbf{(Inner bound 1: DF strategy)} A single-letter characterization of the region $\mathcal{R}^{d1}$
($\mathcal{R}^{d1}\subseteq \mathcal{R}^{d}$) is as follows,
\begin{eqnarray*}
&&\mathcal{R}^{d1}=\{(R_{1}, R_{2}): R_{1}, R_{2}\geq 0,\\
&&R_{1}\leq \min\{I(X_{1};Y_{r}|X_{r},X_{2},V_{1},V_{2}),I(X_{1},X_{r};Y|X_{2},V_{2})\}-I(X_{1};Z),\\
&&R_{2}\leq \min\{I(X_{2};Y_{r}|X_{r},X_{1},V_{1},V_{2}),I(X_{2},X_{r};Y|X_{1},V_{1})\}-I(X_{2};Z),\\
&&R_{1}+R_{2}\leq \min\{I(X_{1},X_{2};Y_{r}|X_{r},V_{1},V_{2}),I(X_{1},X_{2},X_{r};Y)\}-I(X_{1},X_{2};Z)\},
\end{eqnarray*}
for some distribution
\begin{eqnarray*}
&&P_{Y,Z,Y_{r},X_{r},X_{1},X_{2},V_{1},V_{2}}(y,z,y_{r},x_{r},x_{1},x_{2},v_{1},v_{2})=\\
&&P_{Y,Z,Y_{r}|X_{r},X_{1},X_{2}}(y,z,y_{r}|x_{r},x_{1},x_{2})
P_{X_{r}|V_{1},V_{2}}(x_{r}|v_{1},v_{2})P_{X_{1}|V_{1}}(x_{1}|v_{1})P_{X_{2}|V_{2}}(x_{2}|v_{2})P_{V_{1}}(v_{1})P_{V_{2}}(v_{2}).
\end{eqnarray*}
\end{theorem}

\begin{IEEEproof}

The achievable coding scheme is a combination of \cite{SKM, KGG} and \cite{TY1}, and the details about the proof
are provided in  Appendix \ref{appen1}.
\end{IEEEproof}

\begin{remark}\label{R1}
There are some notes on Theorem \ref{T1}, see the following.
\begin{itemize}

\item If we let $Z=const$ (which implies that there is no wiretapper), the region $\mathcal{R}^{d1}$ reduces to the following achievable region $\mathcal{R}^{marc}$, where
\begin{eqnarray}\label{e203}
&&\mathcal{R}^{marc}=\{(R_{1}, R_{2}): R_{1}, R_{2}\geq 0,\nonumber\\
&&R_{1}\leq \min\{I(X_{1};Y_{r}|X_{r},X_{2},V_{1},V_{2}),I(X_{1},X_{r};Y|X_{2},V_{2})\},\nonumber\\
&&R_{2}\leq \min\{I(X_{2};Y_{r}|X_{r},X_{1},V_{1},V_{2}),I(X_{2},X_{r};Y|X_{1},V_{1})\},\nonumber\\
&&R_{1}+R_{2}\leq \min\{I(X_{1},X_{2};Y_{r}|X_{r},V_{1},V_{2}),I(X_{1},X_{2},X_{r};Y)\}\}.
\end{eqnarray}
Here note that the achievable region $\mathcal{R}^{marc}$ is exactly the same as the achievable
DF region (DF inner bound on the capacity region) of the
discrete memoryless multiple-access relay channel \cite{SKM, KGG}.

\item If we let $Y_{r}=Y$ and $V_{1}=V_{2}=X_{r}=const$ (which implies that there is no relay),
the region $\mathcal{R}^{d1}$ reduces to the region $\mathcal{R}^{mac-wt}$, where
\begin{eqnarray}\label{e204}
&&\mathcal{R}^{mac-wt}=\{(R_{1}, R_{2}): R_{1}, R_{2}\geq 0,\nonumber\\
&&R_{1}\leq I(X_{1};Y|X_{2})-I(X_{1};Z),\nonumber\\
&&R_{2}\leq I(X_{2};Y|X_{1})-I(X_{2};Z),\nonumber\\
&&R_{1}+R_{2}\leq I(X_{1},X_{2};Y)-I(X_{1},X_{2};Z)\}.
\end{eqnarray}
Also note that the region $\mathcal{R}^{mac-wt}$ is exactly the same as the achievable secrecy rate region of
discrete memoryless multiple-access wiretap channel \cite{TY1}.

\end{itemize}
\end{remark}

The second step is to characterize the inner bound on the secrecy capacity region $\mathcal{R}^{d}$
by using the noise and forward (NF) strategy. In the
NF Strategy, the relay node does not attempt to decode the messages but sends sequences that are independent of
the transmitters' messages, and these sequences aid in confusing the wiretapper.

More specifically, for a given input distribution of the relay,
if the corresponding mutual information with the
legitimate receiver's output is not less than that with the wiretapper's
output, we allow the legitimate receiver to decode the sequence of the relay, and the wiretapper can
not decode it. Therefore, in this case, the sequence of the relay can
be viewed as a noise signal to confuse the wiretapper.

On the other hand, if the corresponding mutual information with the
legitimate receiver's output is not more than that with the wiretapper's
output, we allow both the receivers to decode the sequence of the relay. In this case,
the sequence of the relay does not make any contribution to the
security of the discrete memoryless MARC-WT.

The following Theorem \ref{T2} shows the NF inner bound on $\mathcal{R}^{d}$.

\begin{theorem}\label{T2}
\textbf{(Inner bound 2: NF strategy)} A single-letter characterization of the region $\mathcal{R}^{d2}$
($\mathcal{R}^{d2}\subseteq \mathcal{R}^{d}$) is as follows,
\begin{eqnarray*}
&&\mathcal{R}^{d2}=\mbox{convex closure of}\quad(\mathcal{L}^{1}\bigcup \mathcal{L}^{2}),
\end{eqnarray*}
where $\mathcal{L}^{1}$ is given by
\begin{eqnarray*}
&&\mathcal{L}^{1}=\bigcup_{\mbox{\tiny$\begin{array}{c}
P_{Y,Z,Y_{r},X_{r},X_{1},X_{2}}\textbf{:}\\
I(X_{r};Y)\geq I(X_{r};Z)\end{array}$}}
\left\{
\begin{array}{ll}
(R_{1}, R_{2}): R_{1}, R_{2}\geq 0,\\
R_{1}\leq I(X_{1};Y|X_{2},X_{r})-I(X_{1},X_{r};Z)+R_{r},\\
R_{2}\leq I(X_{2};Y|X_{1},X_{r})-I(X_{2},X_{r};Z)+R_{r},\\
R_{1}+R_{2}\leq I(X_{1},X_{2};Y|X_{r})-I(X_{1},X_{2},X_{r};Z)+R_{r}.
\end{array}
\right\},
\end{eqnarray*}
$R_{r}$ denotes
\begin{eqnarray*}
&&R_{r}=\min\{I(X_{r};Y), I(X_{r};Z|X_{1}), I(X_{r};Z|X_{2})\},
\end{eqnarray*}
and $\mathcal{L}^{2}$ is given by
\begin{eqnarray*}
&&\mathcal{L}^{2}=\bigcup_{\mbox{\tiny$\begin{array}{c}
P_{Y,Z,Y_{r},X_{r},X_{1},X_{2}}\textbf{:}\\
I(X_{r};Z)\geq I(X_{r};Y)\end{array}$}}
\left\{
\begin{array}{ll}
(R_{1}, R_{2}): R_{1}, R_{2}\geq 0,\\
R_{1}\leq I(X_{1};Y|X_{2},X_{r})-I(X_{1};Z|X_{r}),\\
R_{2}\leq I(X_{2};Y|X_{1},X_{r})-I(X_{2};Z|X_{r}),\\
R_{1}+R_{2}\leq I(X_{1},X_{2};Y|X_{r})-I(X_{1},X_{2};Z|X_{r}).
\end{array}
\right\},
\end{eqnarray*}
here the joint probability $P_{Y,Z,Y_{r},X_{r},X_{1},X_{2}}(y,z,y_{r},x_{r},x_{1},x_{2},u)$ satisfies
\begin{eqnarray*}
&&P_{Y,Z,Y_{r},X_{r},X_{1},X_{2}}(y,z,y_{r},x_{r},x_{1},x_{2})=P_{Y,Z,Y_{r}|X_{r},X_{1},X_{2}}(y,z,y_{r}|x_{r},x_{1},x_{2})
P_{X_{r}}(x_{r})P_{X_{1}}(x_{1})P_{X_{2}}(x_{2}).
\end{eqnarray*}

\end{theorem}

\begin{IEEEproof}

The achievable coding scheme is a combination of \cite[Theorem 3]{LG} and \cite{TY1}, and the details about the proof
are provided in  Appendix \ref{appen2}.
\end{IEEEproof}

\begin{remark}\label{R2}
There are some notes on Theorem \ref{T2}, see the following.
\begin{itemize}

\item Since the two regions $\mathcal{L}^{1}$ and $\mathcal{L}^{2}$ are not necessarily contained by one another, by using time-sharing arguments,
it is easy to find a new achievable region which is the convex-closure of the union of the two
regions.

\item The region $\mathcal{L}^{1}$ is characterized under the condition that
for a given input distribution of the relay, the corresponding mutual information with the
legitimate receiver's output is not less than that with the wiretapper's
output ($I(X_{r};Y)\geq I(X_{r};Z)$). Then, in this case, the legitimate receiver is allowed to decode the sequence of the relay,
and the wiretapper is not allowed to decode it. The rate of the sequence is defined as $R_{r}=\min\{I(X_{r};Y), I(X_{r};Z|X_{1}), I(X_{r};Z|X_{2})\}$, and
the sequence is viewed as pure noise for the wiretapper.

\item The region $\mathcal{L}^{2}$ is characterized under the condition that
for a given input distribution of the relay, the corresponding mutual information with the
legitimate receiver's output is not more than that with the wiretapper's
output ($I(X_{r};Y)\leq I(X_{r};Z)$). Then, in this case, both the legitimate receiver and the wiretapper are allowed to decode the
sequence of the relay. The rate of the sequence is defined as $R_{r}=I(X_{r};Y)$, and
the sequence does not make any contribution to the
security of the discrete memoryless MARC-WT.

\end{itemize}
\end{remark}

The third step is to characterize the inner bound on the
secrecy capacity region $\mathcal{R}^{d}$ by using a combination of Cover-
El Gamal¡¯s compress and forward (CF) strategy \cite{CG} and the NF
strategy provided in Theorem \ref{T2}, i.e., in addition to the independent codewords, the
relay also sends a quantized version of its noisy observations
to the legitimate receiver. This noisy version of the relay's
observations helps the legitimate receiver in decoding the transmitters' messages, while the independent codewords help in confusing
the wiretapper.
The following Theorem \ref{T3} shows the CF inner bound on $\mathcal{R}^{d}$.

\begin{theorem}\label{T3}
\textbf{(Inner bound 3: CF strategy)} A single-letter characterization of the region $\mathcal{R}^{d3}$
($\mathcal{R}^{d3}\subseteq \mathcal{R}^{d}$) is as follows,
\begin{eqnarray*}
&&\mathcal{R}^{d3}=\mbox{convex closure of}\quad(\mathcal{L}^{3}\bigcup \mathcal{L}^{4}),
\end{eqnarray*}
where $\mathcal{L}^{3}$ is given by
\begin{eqnarray*}
&&\mathcal{L}^{3}=\bigcup_{\mbox{\tiny$\begin{array}{c}
P_{Y,Z,Y_{r},\hat{Y}_{r},X_{r},X_{1},X_{2}}:I(X_{r};Y)\geq I(X_{r};Z)\\
R^{*}_{r1}-R^{*}\geq I(Y_{r};\hat{Y}_{r}|X_{r})\end{array}$}}
\left\{
\begin{array}{ll}
(R_{1}, R_{2}): R_{1}, R_{2}\geq 0,\\
R_{1}\leq I(X_{1};Y,\hat{Y}_{r}|X_{2},X_{r})-I(X_{1},X_{r};Z)+R^{*},\\
R_{2}\leq I(X_{2};Y,\hat{Y}_{r}|X_{1},X_{r})-I(X_{2},X_{r};Z)+R^{*},\\
R_{1}+R_{2}\leq I(X_{1},X_{2};Y,\hat{Y}_{r}|X_{r})-I(X_{1},X_{2},X_{r};Z)+R^{*}.
\end{array}
\right\},
\end{eqnarray*}
$R^{*}_{r1}=\min\{I(X_{r};Z|X_{1}), I(X_{r};Z|X_{2}), I(X_{r};Y)\}$,  $R^{*}$ is the rate of pure noise generated by the relay to confuse the wiretapper,
$R^{*}_{r1}-R^{*}$ is the part of the rate allocated to send the compressed signal $\hat{Y}_{r}$ to help
the legitimate receiver,
and $\mathcal{L}^{4}$ is given by
\begin{eqnarray*}
&&\mathcal{L}^{4}=\bigcup_{\mbox{\tiny$\begin{array}{c}
P_{Y,Z,Y_{r},\hat{Y}_{r},X_{r},X_{1},X_{2}}:I(X_{r};Z)\geq I(X_{r};Y)\\
I(X_{r};Y)\geq I(Y_{r};\hat{Y}_{r}|X_{r})\end{array}$}}
\left\{
\begin{array}{ll}
(R_{1}, R_{2}): R_{1}, R_{2}\geq 0,\\
R_{1}\leq I(X_{1};Y,\hat{Y}_{r}|X_{2},X_{r})-I(X_{1};Z|X_{r}),\\
R_{2}\leq I(X_{2};Y,\hat{Y}_{r}|X_{1},X_{r})-I(X_{2};Z|X_{r}),\\
R_{1}+R_{2}\leq I(X_{1},X_{2};Y,\hat{Y}_{r}|X_{r})-I(X_{1},X_{2};Z|X_{r}).
\end{array}
\right\}.
\end{eqnarray*}

The joint probability $P_{Y,Z,Y_{r},\hat{Y}_{r},X_{r},X_{1},X_{2}}(y,z,y_{r},\hat{y}_{r},x_{r},x_{1},x_{2})$ satisfies
\begin{eqnarray*}
&&P_{Y,Z,Y_{r},\hat{Y}_{r},X_{r},X_{1},X_{2}}(y,z,y_{r},\hat{y}_{r},x_{r},x_{1},x_{2})=\\
&&P_{\hat{Y}_{r}|Y_{r},X_{r}}(\hat{y}_{r}|y_{r},x_{r})P_{Y,Z,Y_{r}|X_{r},X_{1},X_{2}}(y,z,y_{r}|x_{r},x_{1},x_{2})
P_{X_{r}}(x_{r})P_{X_{1}}(x_{1})P_{X_{2}}(x_{2}).
\end{eqnarray*}

\end{theorem}

\begin{IEEEproof}

The achievable coding scheme is a combination of \cite[Theorem 4]{LG} and \cite{TY1}, and the details about the proof
are provided in  Appendix \ref{appen3}.
\end{IEEEproof}

\begin{remark}\label{R3}
There are some notes on Theorem \ref{T3}, see the following.
\begin{itemize}

\item Since the two regions $\mathcal{L}^{3}$ and $\mathcal{L}^{4}$ are not necessarily contained by one another, by using time-sharing arguments,
it is easy to find a new achievable region which is the convex-closure of the union of the two
regions.

\item The region $\mathcal{L}^{3}$ is characterized under the condition that for a given input
distribution of the relay, the corresponding mutual information with the
legitimate receiver's output is not less than that with the wiretapper's
output ($I(X_{r};Y)\geq I(X_{r};Z)$). Then, in this case, the legitimate receiver is allowed to
decode the sequence of the relay, and the wiretapper is not allowed to
decode it. Here note that if $R^{*}=R^{*}_{r1}$, this scheme is exactly the same as the NF scheme.

\item The region $\mathcal{L}^{4}$ is characterized under the condition that for a given input
distribution of the relay, the corresponding mutual information with the
legitimate receiver's output is not more than that with the wiretapper's
output ($I(X_{r};Y)\leq I(X_{r};Z)$). Then, in this case, both the legitimate receiver and the wiretapper are allowed to decode the sequence of the relay.
However, the relay can still help to enhance the security of the discrete memoryless MARC-WT
by sending the compressed signal $\hat{Y}_{r}$ to the legitimate receiver.

\end{itemize}
\end{remark}

\subsection{Outer bound on the secrecy capacity region of the degraded discrete memoryless MARC-WT}\label{sec2.2}

Compared with the discrete memoryless MARC-WT (see Figure \ref{f1}), the degraded case implies the existence
of a Markov chain $(X_{1},X_{2},X_{r},Y_{r})\rightarrow Y\rightarrow Z$.
The secrecy capacity region $\mathcal{R}^{dd}$ of the degraded discrete memoryless MARC-WT is a set composed of
all achievable secrecy rate pairs $(R_{1},R_{2})$.
An outer bound
on $\mathcal{R}^{dd}$ is provided in the following Theorem \ref{T4.1}.

\begin{theorem}\label{T4.1}
\textbf{(Outer bound)} A single-letter characterization of the region $\mathcal{R}^{ddo}$
($\mathcal{R}^{dd}\subseteq \mathcal{R}^{ddo}$) is as follows,
\begin{eqnarray*}
&&\mathcal{R}^{ddo}=\{(R_{1}, R_{2}):R_{1}, R_{2}\geq 0,\\
&&R_{1}\leq I(X_{1},X_{r};Y|X_{2},U)-I(X_{1};Z|U)\\
&&R_{2}\leq I(X_{2},X_{r};Y|X_{1},U)-I(X_{2};Z|U)\\
&&R_{1}+R_{2}\leq I(X_{1},X_{2},X_{r};Y|U)-I(X_{1},X_{2};Z|U)\}\\
\end{eqnarray*}
for some distribution
\begin{eqnarray*}
&&P_{Z,Y,Y_{r},X_{r},X_{1},X_{2},U}(z,y,y_{r},x_{r},x_{1},x_{2},u)=\\
&&P_{Z|Y}(z|y)P_{Y,Y_{r}|X_{1},X_{2},X_{r}}(y,y_{r}|x_{1},x_{2},x_{r})
P_{U,X_{1},X_{2},X_{r}}(u,x_{1},x_{2},x_{r}).
\end{eqnarray*}
\end{theorem}

\begin{IEEEproof}

The details about the proof
are provided in  Appendix \ref{appen4}.
\end{IEEEproof}

\begin{remark}\label{R4}

The outer bound on the secrecy capacity region of the degraded discrete memoryless MARC-WT is generally loose,
but it is still useful for the analysis of the outer bound on the secrecy capacity region of the Gaussian MARC-WT,
and this is because the scalar Gaussian MARC-WT is always degraded. The capacity results on the Gaussian MARC-WT
will be given in the next section.

\end{remark}

\section{Gaussian multiple-access relay wiretap channel}\label{secIII}
\setcounter{equation}{0}

In this section, we investigate the Gaussian multiple-access relay wiretap channel (GMARC-WT). The signal received at each node
is given by
\begin{eqnarray}
&&Y_{r}=X_{1}+X_{2}+Z_{r},\nonumber\\
&&Y=X_{1}+X_{2}+X_{r}+Z_{1},\nonumber\\
&&Z=X_{1}+X_{2}+X_{r}+Z_{2},
\end{eqnarray}
where $Z_{r}\sim \mathcal{N}(0, N_{r})$, $Z_{1}\sim \mathcal{N}(0, N_{1})$, $Z_{2}\sim \mathcal{N}(0, N_{2})$,
and they are independent. The Gaussian noise vectors $Z^{N}_{r}$, $Z^{N}_{1}$ and $Z^{N}_{2}$ are composed of i.i.d. components with
probability distributions $Z_{r}\sim \mathcal{N}(0, N_{r})$, $Z_{1}\sim \mathcal{N}(0, N_{1})$ and $Z_{2}\sim \mathcal{N}(0, N_{2})$, respectively.
The average power
constraints of $X_{1}^{N}$, $X_{2}^{N}$ and $X_{r}^{N}$ are  $\frac{1}{N}\sum_{i=1}^{N}E[X_{1,i}^{2}]\leq P_{1}$,
$\frac{1}{N}\sum_{i=1}^{N}E[X_{2,i}^{2}]\leq P_{2}$ and $\frac{1}{N}\sum_{i=1}^{N}E[X_{r,i}^{2}]\leq P_{r}$, respectively.

The remainder of this section is organized as follows. Subsection \ref{secIII.1} shows the achievable secrecy
rate regions of GMARC-WT, and the numerical examples and discussions are given in Subsection \ref{secIII.2}.

\subsection{Capacity results on GMARC-WT}\label{secIII.1}

\begin{theorem}\label{T4}
The DF inner bound on the secrecy capacity region of the GMARC-WT is given by
\begin{equation}\label{e402}
\mathcal{R}^{g1}=\bigcup_{0\leq \gamma\leq 1}
\left\{
\begin{array}{ll}
(R_{1}, R_{2}): R_{1}, R_{2}\geq 0,\\
R_{1}\leq \min\{\frac{1}{2}\log(1+\frac{P_{1}}{N_{r}}), \frac{1}{2}\log(1+\frac{P_{1}+\gamma P_{r}}{N_{1}})\}-\frac{1}{2}\log\frac{P_{1}+P_{2}+P_{r}+N_{2}}{P_{2}+P_{r}+N_{2}},\\
R_{2}\leq \min\{\frac{1}{2}\log(1+\frac{P_{2}}{N_{r}}), \frac{1}{2}\log(1+\frac{P_{2}+(1-\gamma) P_{r}}{N_{1}})\}-\frac{1}{2}\log\frac{P_{1}+P_{2}+P_{r}+N_{2}}{P_{1}+P_{r}+N_{2}},\\
R_{1}+R_{2}\leq \min\{\frac{1}{2}\log(1+\frac{P_{1}+P_{2}}{N_{r}}), \frac{1}{2}\log(1+\frac{P_{1}+P_{2}+P_{r}}{N_{1}})\}-\frac{1}{2}\log\frac{P_{1}+P_{2}+P_{r}+N_{2}}{P_{r}+N_{2}}.
\end{array}
\right\}.
\end{equation}
\end{theorem}

\begin{IEEEproof}

First, let $X_{r}=V_{1}+V_{2}$, where $V_{1}\sim \mathcal{N}(0, \gamma P_{r})$ and
$V_{2}\sim \mathcal{N}(0, (1-\gamma) P_{r})$.

Let $X_{1}=\sqrt{\frac{(1-\alpha)P_{1}}{\gamma P_{r}}}V_{1}+X_{10}$, where $0\leq \alpha\leq 1$ and
$X_{10}\sim \mathcal{N}(0, \alpha P_{1})$.

Analogously, let $X_{2}=\sqrt{\frac{(1-\beta)P_{2}}{(1-\gamma) P_{r}}}V_{2}+X_{20}$, where $0\leq \beta\leq 1$ and
$X_{20}\sim \mathcal{N}(0, \beta P_{2})$.

Here note that $V_{1}$, $V_{2}$, $X_{10}$ and $X_{20}$ are independent random variables.

The region $\mathcal{R}^{g1}$ is obtained by substituting the above definitions into Theorem \ref{T1}, and maximizing $\alpha$ and $\beta$
(the maximum of $\mathcal{R}^{g1}$ is achieved when $\alpha=\beta=1$). Thus, the proof of Theorem \ref{T4} is completed.

\end{IEEEproof}

\begin{theorem}\label{T5}
The NF inner bound on the secrecy capacity region of the GMARC-WT is given by
\begin{eqnarray*}
&&\mathcal{R}^{g2}=\mbox{convex closure of}\quad(\mathcal{G}^{1}\bigcup \mathcal{G}^{2}),
\end{eqnarray*}
where $\mathcal{G}^{1}$ is given by
\begin{eqnarray*}
&&\mathcal{G}^{1}=\bigcup_{N_{1}\leq N_{2}}
\left\{
\begin{array}{ll}
(R_{1}, R_{2}): R_{1}, R_{2}\geq 0,\\
R_{1}\leq \frac{1}{2}\log(1+\frac{P_{1}}{N_{1}})-\frac{1}{2}\log(1+\frac{P_{1}+P_{r}}{P_{2}+N_{2}})+R_{r},\\
R_{2}\leq \frac{1}{2}\log(1+\frac{P_{2}}{N_{1}})-\frac{1}{2}\log(1+\frac{P_{2}+P_{r}}{P_{1}+N_{2}})+R_{r},\\
R_{1}+R_{2}\leq \frac{1}{2}\log(1+\frac{P_{1}+P_{2}}{N_{1}})-\frac{1}{2}\log(1+\frac{P_{1}+P_{2}+P_{r}}{N_{2}})+R_{r}.
\end{array}
\right\},
\end{eqnarray*}
$R_{r}=\min\{\frac{1}{2}\log(1+\frac{P_{r}}{P_{1}+P_{2}+N_{1}}), \frac{1}{2}\log(1+\frac{P_{r}}{P_{2}+N_{2}}), \frac{1}{2}\log(1+\frac{P_{r}}{P_{1}+N_{2}})\}$,
and $\mathcal{G}^{2}$ is given by
\begin{eqnarray*}
&&\mathcal{G}^{2}=\bigcup_{N_{1}\geq N_{2}}
\left\{
\begin{array}{ll}
(R_{1}, R_{2}): R_{1}, R_{2}\geq 0,\\
R_{1}\leq \frac{1}{2}\log(1+\frac{P_{1}}{N_{1}})-\frac{1}{2}\log(1+\frac{P_{1}}{P_{2}+N_{2}}),\\
R_{2}\leq \frac{1}{2}\log(1+\frac{P_{2}}{N_{1}})-\frac{1}{2}\log(1+\frac{P_{2}}{P_{1}+N_{2}}),\\
R_{1}+R_{2}\leq \frac{1}{2}\log(1+\frac{P_{1}+P_{2}}{N_{1}})-\frac{1}{2}\log(1+\frac{P_{1}+P_{2}}{N_{2}}).
\end{array}
\right\}.
\end{eqnarray*}

\end{theorem}

\begin{IEEEproof}

Here note that $N_{1}\leq N_{2}$ implies $I(X_{r};Y)\geq I(X_{r};Z)$. The region $\mathcal{G}^{1}$
is obtained by substituting $X_{1}\sim \mathcal{N}(0, P_{1})$, $X_{2}\sim \mathcal{N}(0, P_{2})$ and $X_{r}\sim \mathcal{N}(0, P_{r})$
into the region $\mathcal{L}^{1}$ of Theorem \ref{T2}, and using the fact that
$X_{1}$, $X_{2}$ and $X_{r}$ are independent random variables.

Analogously, $N_{1}\geq N_{2}$ implies $I(X_{r};Y)\leq I(X_{r};Z)$. The region $\mathcal{G}^{2}$
is obtained by substituting $X_{1}\sim \mathcal{N}(0, P_{1})$, $X_{2}\sim \mathcal{N}(0, P_{2})$ and $X_{r}\sim \mathcal{N}(0, P_{r})$
into the region $\mathcal{L}^{2}$ of Theorem \ref{T2}, and using the fact that
$X_{1}$, $X_{2}$ and $X_{r}$ are independent random variables.
Thus, the proof of Theorem \ref{T5} is completed.

\end{IEEEproof}

\begin{theorem}\label{T6}

The CF inner bound on the secrecy capacity region of the GMARC-WT is given by
\begin{eqnarray*}
&&\mathcal{R}^{g3}=\mbox{convex closure of}\quad(\mathcal{G}^{3}\bigcup \mathcal{G}^{4}),
\end{eqnarray*}
where $\mathcal{G}^{3}$ is given by
\begin{eqnarray*}
&&\mathcal{G}^{3}=\bigcup_{N_{1}\leq N_{2}}
\left\{
\begin{array}{ll}
(R_{1}, R_{2}): R_{1}, R_{2}\geq 0,\\
R_{1}\leq \frac{1}{2}\log(1+\frac{P_{1}(Q+N_{1}+N_{r})}{N_{1}(N_{r}+Q)})-\frac{1}{2}\log(1+\frac{P_{1}+P_{r}}{P_{2}+N_{2}})+R^{*},\\
R_{2}\leq \frac{1}{2}\log(1+\frac{P_{2}(Q+N_{1}+N_{r})}{N_{1}(N_{r}+Q)})-\frac{1}{2}\log(1+\frac{P_{2}+P_{r}}{P_{1}+N_{2}})+R^{*},\\
R_{1}+R_{2}\leq \frac{1}{2}\log(1+\frac{(P_{1}+P_{2})(Q+N_{1}+N_{r})}{N_{1}(N_{r}+Q)})-\frac{1}{2}\log(1+\frac{P_{1}+P_{2}+P_{r}}{N_{2}})+R^{*}.
\end{array}
\right\},
\end{eqnarray*}
$Q$ satisfies
\begin{eqnarray*}
\frac{1}{2}\log(1+\frac{P_{1}+P_{2}+N_{r}}{Q})&\leq& \min\{\frac{1}{2}\log(1+\frac{P_{r}}{P_{1}+P_{2}+N_{1}}), \frac{1}{2}\log(1+\frac{P_{r}}{P_{2}+N_{2}}), \frac{1}{2}\log(1+\frac{P_{r}}{P_{1}+N_{2}})\},
\end{eqnarray*}
and $R^{*}$ satisfies
\begin{eqnarray*}
0\leq R^{*}&\leq&\min\{\frac{1}{2}\log(1+\frac{P_{r}}{P_{1}+P_{2}+N_{1}}), \frac{1}{2}\log(1+\frac{P_{r}}{P_{2}+N_{2}}), \frac{1}{2}\log(1+\frac{P_{r}}{P_{1}+N_{2}})\}\\
&&-\frac{1}{2}\log(1+\frac{P_{1}+P_{2}+N_{r}}{Q}),
\end{eqnarray*}
and $\mathcal{G}^{4}$ is given by
\begin{eqnarray*}
&&\mathcal{G}^{4}=\bigcup_{N_{1}\geq N_{2}}
\left\{
\begin{array}{ll}
(R_{1}, R_{2}): R_{1}, R_{2}\geq 0,\\
R_{1}\leq \frac{1}{2}\log(1+\frac{P_{1}(Q+N_{1}+N_{r})}{N_{1}(N_{r}+Q)})-\frac{1}{2}\log(1+\frac{P_{1}}{P_{2}+N_{2}}),\\
R_{2}\leq \frac{1}{2}\log(1+\frac{P_{2}(Q+N_{1}+N_{r})}{N_{1}(N_{r}+Q)})-\frac{1}{2}\log(1+\frac{P_{2}}{P_{1}+N_{2}}),\\
R_{1}+R_{2}\leq \frac{1}{2}\log(1+\frac{(P_{1}+P_{2})(Q+N_{1}+N_{r})}{N_{1}(N_{r}+Q)})-\frac{1}{2}\log(1+\frac{P_{1}+P_{2}}{N_{2}}).
\end{array}
\right\},
\end{eqnarray*}
here $Q$ satisfies
$Q\geq \frac{(P_{1}+P_{2})^{2}+(P_{1}+P_{2})(N_{r}+N_{1})+N_{r}N_{1}}{P_{r}}$.

\end{theorem}

\begin{IEEEproof}

Here note that $N_{1}\leq N_{2}$ implies $I(X_{r};Y)\geq I(X_{r};Z)$. The region $\mathcal{G}^{3}$
is obtained by substituting $X_{1}\sim \mathcal{N}(0, P_{1})$, $X_{2}\sim \mathcal{N}(0, P_{2})$, $X_{r}\sim \mathcal{N}(0, P_{r})$,
$\hat{Y}_{r}=Y_{r}+Z_{Q}$\footnotemark[1] and $Z_{Q}\sim \mathcal{N}(0, Q)$
into the region $\mathcal{L}^{3}$ of Theorem \ref{T3}, and using the fact that
$X_{1}$, $X_{2}$ and $X_{r}$ are independent random variables.

\footnotetext[1]{Here note that $\hat{Y}_{r}=Y_{r}+Z_{Q}$
is from the similar argument for the CF strategy of the Gaussian relay channel \cite[pp. 402-403]{Gamal}.}

Analogously, $N_{1}\geq N_{2}$ implies $I(X_{r};Y)\leq I(X_{r};Z)$. The region $\mathcal{G}^{4}$
is obtained by substituting $X_{1}\sim \mathcal{N}(0, P_{1})$, $X_{2}\sim \mathcal{N}(0, P_{2})$, $X_{r}\sim \mathcal{N}(0, P_{r})$,
$\hat{Y}_{r}=Y_{r}+Z_{Q}$ and $Z_{Q}\sim \mathcal{N}(0, Q)$
into the region $\mathcal{L}^{4}$ of Theorem \ref{T3}, and using the fact that
$X_{1}$, $X_{2}$ and $X_{r}$ are independent random variables.
Thus, the proof of Theorem \ref{T6} is completed.

\end{IEEEproof}

By using Theorem \ref{T4.1}, we provide an outer bound on the secrecy capacity region of the GMARC-WT under the condition that $N_{2}\geq N_{1}$,
see the followings.

\begin{theorem}\label{T6.5}

For the case that $N_{2}\geq N_{1}$, an outer bound $\mathcal{R}^{gout}$ on the secrecy capacity region of the GMARC-WT is given by
\begin{eqnarray*}
&&\mathcal{R}^{gout}=\bigcup_{0\leq \alpha, \beta_{1}, \beta_{2}, \gamma\leq 1}
\left\{
\begin{array}{ll}
(R_{1}, R_{2}): R_{1}\geq 0, R_{2}\geq 0,\\
R_{1}\leq \frac{1}{2}\log(1+\frac{P_{r}(\alpha+\beta_{2}-\alpha\beta_{2})+\beta_{2}P_{1}}{N_{1}})
-\frac{1}{2}\log(\frac{C+\gamma(P_{1}+P_{2}+P_{r}+N_{2}-C)}{N_{2}+P_{r}(\alpha+\beta_{1}-\alpha\beta_{1})+\beta_{1}P_{2}}),\\
R_{2}\leq \frac{1}{2}\log(1+\frac{P_{r}(\alpha+\beta_{1}-\alpha\beta_{1})+\beta_{1}P_{2}}{N_{1}})
-\frac{1}{2}\log(\frac{C+\gamma(P_{1}+P_{2}+P_{r}+N_{2}-C)}{N_{2}+P_{r}(\alpha+\beta_{2}-\alpha\beta_{2})+\beta_{2}P_{1}}),\\
R_{1}+R_{2}\leq \frac{1}{2}\log(\frac{C+\gamma(P_{1}+P_{2}+P_{r}+N_{1}-C)}{N_{1}})
-\frac{1}{2}\log(\frac{C+\gamma(P_{1}+P_{2}+P_{r}+N_{2}-C)}{N_{2}+\alpha P_{r}}),
\end{array}
\right\},
\end{eqnarray*}
where $C$ satisfies
\begin{eqnarray*}
C&=&\max\{N_{2}+P_{r}(\alpha+\beta_{1}-\alpha\beta_{1})+\beta_{1}P_{2}, N_{2}+P_{r}(\alpha+\beta_{2}-\alpha\beta_{2})+\beta_{2}P_{1}\}.
\end{eqnarray*}

\end{theorem}

\begin{IEEEproof}

See Appendix \ref{appen5}.

\end{IEEEproof}

\begin{theorem}\label{T7}

Finally, remember that \cite{TY1} provides an achievable secrecy rate region $\mathcal{R}^{Gi}$ of the Gaussian multiple-access wiretap channel (GMAC-WT),
and it is given by
\begin{eqnarray*}
&&\mathcal{R}^{Gi}=
\left\{
\begin{array}{ll}
(R_{1}, R_{2}): R_{1}, R_{2}\geq 0,\\
R_{1}\leq \frac{1}{2}\log(1+\frac{P_{1}}{N_{1}})-\frac{1}{2}\log(1+\frac{P_{1}}{N_{2}+P_{2}})\\
R_{2}\leq \frac{1}{2}\log(1+\frac{P_{2}}{N_{1}})-\frac{1}{2}\log(1+\frac{P_{2}}{N_{2}+P_{1}})\\
R_{1}+R_{2}\leq \frac{1}{2}\log(1+\frac{P_{1}+P_{2}}{N_{1}})-\frac{1}{2}\log(1+\frac{P_{1}+P_{2}}{N_{2}})
\end{array}
\right\}.
\end{eqnarray*}

\end{theorem}

\begin{IEEEproof}

The proof is in \cite{TY1}, and it is omitted here.

\end{IEEEproof}

\subsection{Numerical Examples and Discussions}\label{secIII.2}

Letting $P_{1}=5$, $P_{2}=6$, $P_{r}=20$, $N_{1}=2$, $N_{2}=14$ and $Q=200$,
the following Figure \ref{f2}, \ref{f3}, \ref{f4} and \ref{f5} show the
inner and outer bounds on the secrecy capacity region of the GMARC-WT for different values of $N_{r}$.

Compared with the achievable secrecy rate region $\mathcal{R}^{Gi}$ of GMAC-WT,
it is easy to see that the NF region ($\mathcal{R}^{g2}$) and the CF region ($\mathcal{R}^{g3}$) enhance the region $\mathcal{R}^{Gi}$ (no relay).
The CF region is always smaller than the NF region, and
when $Q\rightarrow \infty$, the CF region tends to the NF region.
For the DF region ($\mathcal{R}^{g1}$), we find that when $N_{r}$ is much larger than $N_{1}$, $\mathcal{R}^{g1}$
is even smaller than $\mathcal{R}^{Gi}$ (see Figure \ref{f2}). When $N_{r}$ is close to $N_{1}$ (still larger than $N_{1}$),
$\mathcal{R}^{g1}$ is larger than $\mathcal{R}^{Gi}$, but it is still smaller than the NF and CF regions (see Figure \ref{f3}).
When $N_{r}$ is smaller than $N_{1}$, as we can see in Figure \ref{f4} and \ref{f5}, the DF region $\mathcal{R}^{g1}$ is larger than the NF and CF regions.

Figure \ref{f2}, \ref{f3}, \ref{f4} and \ref{f5} also show that there exists a gap between the
inner and outer bounds, and the gap is reduced as $N_{r}$ decreases.

\begin{figure}[htb]
\centering
\includegraphics[scale=0.45]{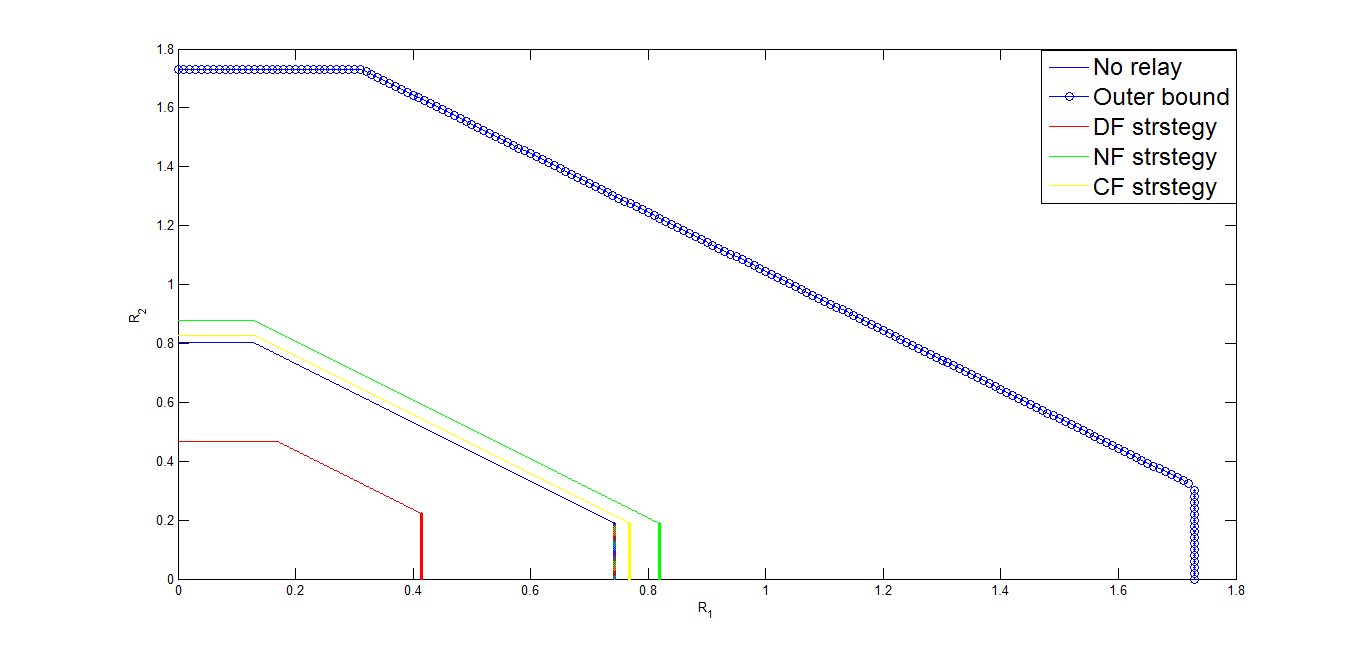}
\caption{The bounds on the secrecy capacity region of GMARC-WT for $N_{r}=5$}
\label{f2}
\end{figure}

\begin{figure}[htb]
\centering
\includegraphics[scale=0.45]{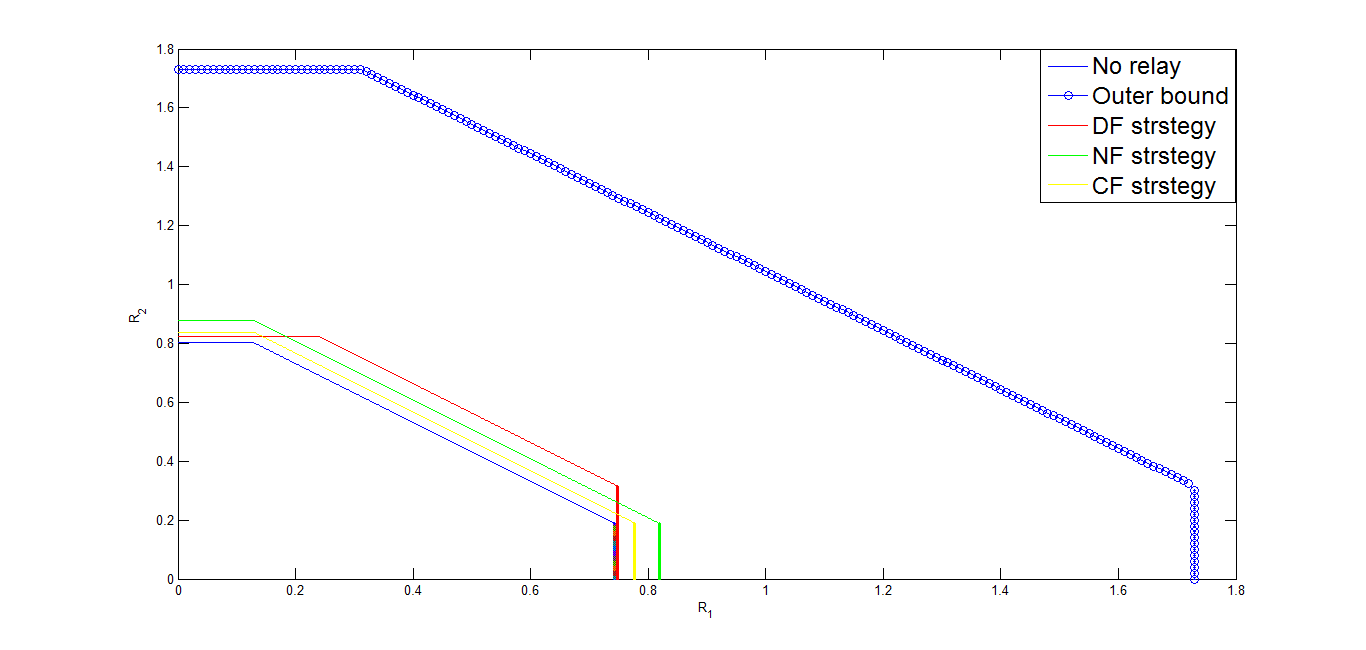}
\caption{The bounds on the secrecy capacity region of GMARC-WT for $N_{r}=2.3$}
\label{f3}
\end{figure}

\begin{figure}[htb]
\centering
\includegraphics[scale=0.45]{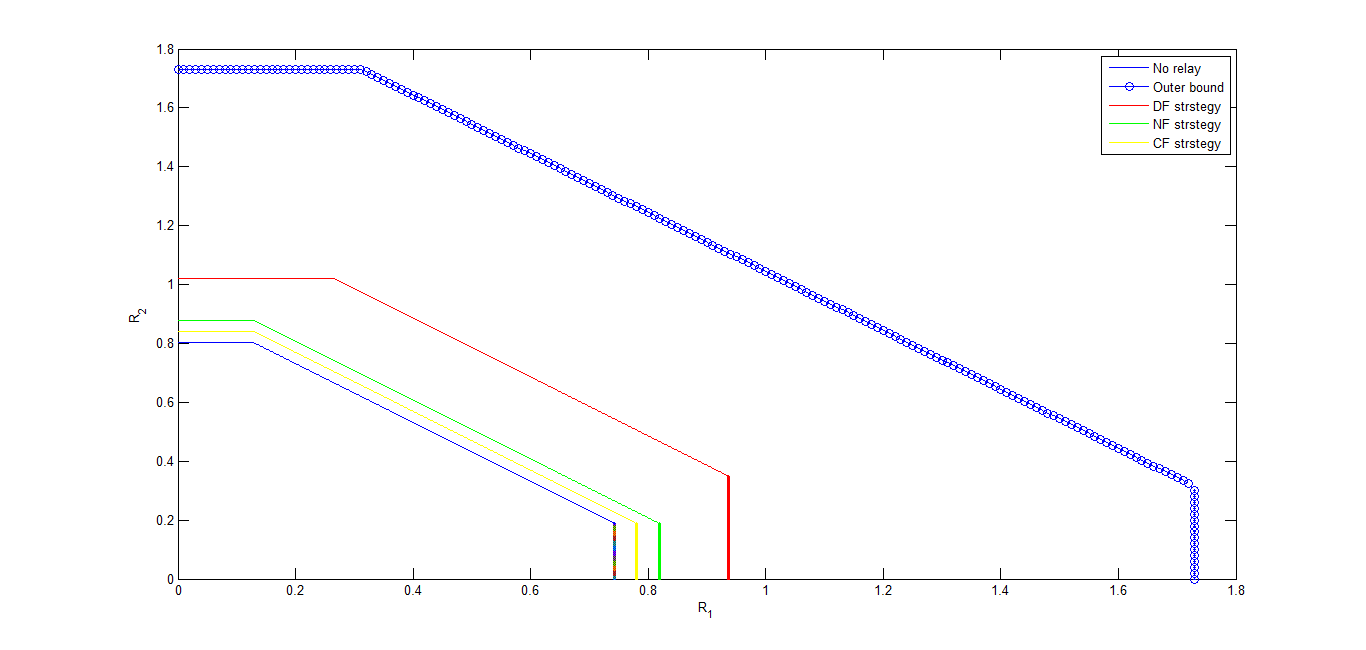}
\caption{The bounds on the secrecy capacity region of GMARC-WT for $N_{r}=1.6$}
\label{f4}
\end{figure}

\begin{figure}[htb]
\centering
\includegraphics[scale=0.45]{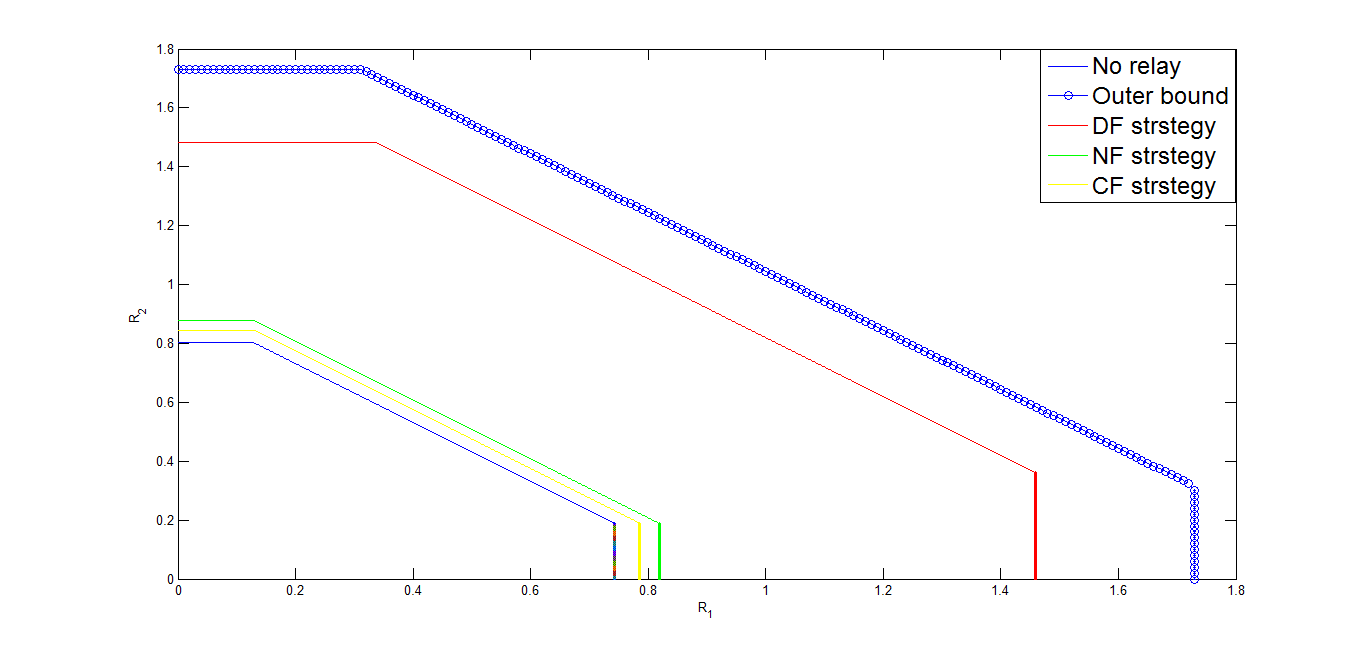}
\caption{The bounds on the secrecy capacity region of GMARC-WT for $N_{r}=0$}
\label{f5}
\end{figure}

\section{Conclusion}\label{secV}

In this paper, first, we provide three inner bounds on the
secrecy capacity region (achievable secrecy rate regions) of
the discrete memoryless model of Figure \ref{f1}. The decode-forward (DF), noise-forward
(NF), and compress-forward (CF) relay strategies are used in
the construction of these inner bounds.
Second, we investigate the degraded discrete memoryless MARC-WT, and present an outer bound on the secrecy capacity region of this degraded case.
Finally,  we study the Gaussian MARC-WT, and
find that the NF and CF strategies help to enhance Tekin-Yener's achievable
secrecy rate region of Gaussian MAC-WT. Moreover, we find that if the channel from the
transmitters to the relay is less noisy than the channels
from the transmitters to the legitimate receiver and the wiretapper, the achievable
secrecy rate region of the DF
strategy is even larger than the corresponding regions of the NF and CF strategies.

\section*{Acknowledgement}

The authors would like to thank Professor Ning Cai
for his valuable suggestions to improve this paper.
This work was supported by a
sub-project in National Basic Research Program of China under Grant 2012CB316100 on Broadband Mobile Communications at High Speeds,
the National Natural Science Foundation of China under Grant 61301121, and the Fundamental Research Funds for the Central Universities
under Grant 2682014CX099.

\renewcommand{\theequation}{A\arabic{equation}}
\appendices\section{Proof of Theorem \ref{T1}\label{appen1}}
\setcounter{equation}{0}

In order to prove Theorem \ref{T1}, we need to show that any pair $(R_{1}, R_{2})\in \mathcal{R}^{d1}$ is achievable, i.e.,
for any $\epsilon>0$, there exists a sequence of codes $(2^{NR_{1}}, 2^{NR_{2}}, N)$ such that
$\frac{\log\parallel \mathcal{W}_{1}\parallel}{N}=R_{1}$, $\frac{\log\parallel \mathcal{W}_{2}\parallel}{N}=R_{2}$,
$P_{e}\leq \epsilon$ and $\frac{1}{N}H(W_{1},W_{2}|Z^{N})\geq R_{1}+R_{2}-\epsilon$.
The details are as follows.

The coding scheme combines the decode-and-forward (DF) strategy of MARC \cite{KGG}, random binning, superposition coding,
and block Markov coding techniques, see the followings.

Fix the joint probability mass function
$P_{Y,Z,Y_{r}|X_{r},X_{1},X_{2}}(y,z,y_{r}|x_{r},x_{1},x_{2})
P_{X_{r}|V_{1},V_{2}}(x_{r}|v_{1},v_{2})\\P_{X_{1}|V_{1}}(x_{1}|v_{1})P_{X_{2}|V_{2}}(x_{2}|v_{2})P_{V_{1}}(v_{1})P_{V_{2}}(v_{2})$.
For a given $(R_{1}, R_{2})\in \mathcal{R}^{d1}$, define the messages $W_{1}$ and $W_{2}$ taking values in the alphabets
$\mathcal{W}_{1}$ and $\mathcal{W}_{2}$, respectively, where
$$\mathcal{W}_{1}=\{1,2,...,2^{NR_{1}}\}, \,\, \mathcal{W}_{2}=\{1,2,...,2^{NR_{2}}\}.$$

\textbf{Relay Code-books Construction:}

For a given $R^{*}_{1}\geq 0$,
generate at random $2^{N(R_{1}+R^{*}_{1})}$ i.i.d. sequences $v_{1}^{N}$ according to
$P_{V_{1}^{N}}(v_{1}^{N})=\prod_{i=1}^{N}P_{V_{1}}(v_{1,i})$. Index them as $v_{1}^{N}(a_{1},b_{1})$, where
$a_{1}\in \{1,2,...,2^{NR_{1}}\}$ and $b_{1}\in \{1,2,...,2^{NR^{*}_{1}}\}$. For convenience, define $s_{1}=(a_{1},b_{1})$,
where $s_{1}\in \{1,2,...,2^{N(R_{1}+R^{*}_{1})}\}$.

Analogously, for a given $R^{*}_{2}\geq 0$, generate at random $2^{N(R_{2}+R^{*}_{2})}$ i.i.d. sequences $v_{2}^{N}$ according to
$P_{V_{2}^{N}}(v_{2}^{N})=\prod_{i=1}^{N}P_{V_{2}}(v_{2,i})$. Index them as $v_{2}^{N}(a_{2},b_{2})$, where
$a_{2}\in \{1,2,...,2^{NR_{2}}\}$ and $b_{2}\in \{1,2,...,2^{NR^{*}_{2}}\}$. For convenience, define $s_{2}=(a_{2},b_{2})$,
where $s_{2}\in \{1,2,...,2^{N(R_{2}+R^{*}_{2})}\}$.

Generate at random $2^{N(R_{1}+R^{*}_{1}+R_{2}+R^{*}_{2})}$ i.i.d. sequences $x_{r}^{N}$ according to
$P_{X_{r}^{N}|V_{1}^{N},V_{2}^{N}}(x_{r}^{N}|v_{1}^{N},v_{2}^{N})=\\ \prod_{i=1}^{N}P_{X_{r,i}|V_{1,i},V_{2,i}}(x_{r,i}|v_{1,i},v_{2,i})$.
Index them as $x_{r}^{N}(s_{1},s_{2})$, where $s_{1}\in \{1,2,...,2^{N(R_{1}+R^{*}_{1})}\}$ and $s_{2}\in \{1,2,...,2^{N(R_{2}+R^{*}_{2})}\}$.

\textbf{Transmitters' Code-books Construction:}

\begin{itemize}

\item For a given $v_{1}^{N}(s_{1})$, generate at random $2^{N(R_{1}+R^{*}_{1})}$ i.i.d. sequences $x_{1}^{N}(w_{1},w^{*}_{1}|s_{1})$
($w_{1}\in \{1,2,...,2^{NR_{1}}\}, w^{*}_{1}\in \{1,2,...,2^{NR^{*}_{1}}\}, s_{1}\in \{1,2,...,2^{N(R_{1}+R^{*}_{1})}\}$)
according to $\prod_{i=1}^{N}P_{X_{1}|V_{1}}(x_{1,i}|v_{1,i})$.

\item Analogously, for a given $v_{2}^{N}(s_{2})$, generate at random $2^{N(R_{2}+R^{*}_{2})}$ i.i.d. sequences $x_{2}^{N}(w_{2},w^{*}_{2}|s_{2})$
($w_{2}\in \{1,2,...,2^{NR_{2}}\}, w^{*}_{2}\in \{1,2,...,2^{NR^{*}_{2}}\}, s_{2}\in \{1,2,...,2^{N(R_{2}+R^{*}_{2})}\}$)
according to $\prod_{i=1}^{N}P_{X_{2}|V_{2}}(x_{2,i}|v_{2,i})$.
\end{itemize}

\textbf{Encoding:} We exploit the block Markov coding scheme, because, as argued in \cite{CG}, the loss induced by
this scheme is negligible as the number of blocks $n\rightarrow \infty$. For block $i$ ($1\leq i\leq n$), encoding proceeds as follows.

First, for convenience, the messages $w_{1}$, $w^{*}_{1}$, $w_{2}$, $w^{*}_{2}$, $s_{1}$ and $s_{2}$ transmitted in the $i$-th block
are denoted by $w_{1,i}$, $w^{*}_{1,i}$, $w_{2,i}$, $w^{*}_{2,i}$, $s_{1,i}$ and $s_{2,i}$, respectively.

\begin{itemize}

\item \textbf{(Channel encoders)}

1) The message $w^{*}_{1,i}$ ($1\leq i\leq n-1$) is randomly chosen from the set $\{1,2,...,2^{NR^{*}_{1}}\}$.
The transmitter 1 (encoder 1) sends $x_{1}^{N}(w_{1,1},w^{*}_{1,1}|1,1)$ at the first block,
$x_{1}^{N}(w_{1,i},w^{*}_{1,i}|w_{1,i-1},w^{*}_{1,i-1})$ (note that here $s_{1,i}=(w_{1,i-1},w^{*}_{1,i-1})$) from block $2\leq i\leq n-1$,
and $x_{1}^{N}(1,1|w_{1,n-1},w^{*}_{1,n-1})$ at block $n$ ($s_{1,n}=(w_{1,n-1},w^{*}_{1,n-1})$).

2)  The message $w^{*}_{2,i}$ ($1\leq i\leq n-1$) is randomly chosen from the set $\{1,2,...,2^{NR^{*}_{2}}\}$.
The transmitter 2 (encoder 2) sends $x_{2}^{N}(w_{2,1},w^{*}_{2,1}|1,1)$ at the first block,
$x_{2}^{N}(w_{2,i},w^{*}_{2,i}|w_{2,i-1},w^{*}_{2,i-1})$ ($s_{2,i}=(w_{2,i-1},w^{*}_{2,i-1})$) from block $2\leq i\leq n-1$,
and $x_{2}^{N}(1,1|w_{2,n-1},w^{*}_{2,n-1})$ at block $n$ ($s_{2,n}=(w_{2,n-1},w^{*}_{2,n-1})$).

\item \textbf{(Relay encoder)}

The relay sends $(v_{1}^{N}(1,1),v_{2}^{N}(1,1),x_{r}^{N}(1,1,1,1))$ at the first block, and \\
$(v_{1}^{N}(\hat{s}_{1,i}),v_{2}^{N}(\hat{s}_{2,i}),x_{r}^{N}(\hat{s}_{1,i},\hat{s}_{2,i}))$ from block $2\leq i\leq n$, where
$\hat{s}_{1,i}=(\hat{w}_{1,i-1}, \hat{w}^{*}_{1,i-1})$ and $\hat{s}_{2,i}=(\hat{w}_{2,i-1}, \hat{w}^{*}_{2,i-1})$.

\end{itemize}

\textbf{Decoding:} Decoding proceeds as follows.

1) (At the relay) At the end of block $i$ ($1\leq i\leq n$), the relay already has an estimation of the $s_{1,i}$ and $s_{2,i}$
(denoted by $\hat{s}_{1,i}$ and $\hat{s}_{2,i}$, respectively), and will declare that it receives $\hat{w}_{1,i}$, $\hat{w}^{*}_{1,i}$,
$\hat{w}_{2,i}$ and $\hat{w}^{*}_{2,i}$ if this is the only quadruple such that $(x_{1}^{N}(\hat{w}_{1,i},\hat{w}^{*}_{1,i}|\hat{s}_{1,i}),x_{2}^{N}(\hat{w}_{2,i},\hat{w}^{*}_{2,i}|\hat{s}_{2,i}),
x_{r}^{N}(\hat{s}_{1,i},\hat{s}_{2,i}), v_{1}(\hat{s}_{1,i}), v_{2}(\hat{s}_{2,i}), y_{r}^{N}(i))$ are jointly typical.
Here note that $y_{r}^{N}(i)$ indicates the output sequence $y_{r}^{N}$ in block $i$, $\hat{s}_{1,i+1}=(\hat{w}_{1,i}, \hat{w}^{*}_{1,i})$
and $\hat{s}_{2,i+1}=(\hat{w}_{2,i}, \hat{w}^{*}_{2,i})$. The indexes $\hat{s}_{1,i+1}$ and $\hat{s}_{2,i+1}$ will be used in the $i+1$-th block.

Based on the AEP, the error probability $Pr\{(\hat{s}_{1,i+1},\hat{s}_{2,i+1})\neq (s_{1,i+1},s_{2,i+1})\}$
goes to $0$ if
\begin{equation}\label{appen1.7.1}
R_{1}+R^{*}_{1}\leq I(X_{1};Y_{r}|X_{r},V_{1},V_{2},X_{2}),
\end{equation}
\begin{equation}\label{appen1.7.2}
R_{2}+R^{*}_{2}\leq I(X_{2};Y_{r}|X_{r},V_{1},V_{2},X_{1}),
\end{equation}
\begin{equation}\label{appen1.7.3}
R_{1}+R^{*}_{1}+R_{2}+R^{*}_{2}\leq I(X_{1},X_{2};Y_{r}|X_{r},V_{1},V_{2}).
\end{equation}

2) (At the legitimate receiver) The legitimate receiver
decodes from the last block, i.e., block $n$. At the end of block $i+1$, the legitimate receiver already has an estimation of the
$\check{w}_{1,i+1}$, $\check{w}^{*}_{1,i+1}$, $\check{w}_{2,i+1}$ and $\check{w}^{*}_{2,i+1}$, and will declare that it receives
$\check{s}_{1,i+1}$ and $\check{s}_{2,i+1}$ if this is the only pair such that
$(x_{1}^{N}(\check{w}_{1,i+1},\check{w}^{*}_{1,i+1}|\check{s}_{1,i+1}),x_{2}^{N}(\check{w}_{2,i+1},\\ \check{w}^{*}_{2,i+1}|\check{s}_{2,i+1}),
x_{r}^{N}(\check{s}_{1,i+1},\check{s}_{2,i+1}), v_{1}(\check{s}_{1,i+1}), v_{2}(\check{s}_{2,i+1}), y^{N}(i+1))$ are jointly typical.
Here note that $y^{N}(i+1)$ indicates the output sequence $y^{N}$ in block $i+1$, $\check{s}_{1,i+1}=(\check{w}_{1,i}, \check{w}^{*}_{1,i})$
and $\check{s}_{2,i+1}=(\check{w}_{2,i}, \check{w}^{*}_{2,i})$.

Based on the AEP, the error probability $Pr\{(\check{s}_{1,i+1},\check{s}_{2,i+1})\neq (s_{1,i+1},s_{2,i+1})\}$
goes to $0$ if
\begin{equation}\label{appen1.10.1}
R_{1}+R^{*}_{1}\leq I(V_{1},X_{r},X_{1};Y|V_{2},X_{2})\stackrel{(a)}=I(X_{r},X_{1};Y|X_{2},V_{2}),
\end{equation}
\begin{equation}\label{appen1.10.2}
R_{2}+R^{*}_{2}\leq I(V_{2},X_{r},X_{2};Y|V_{1},X_{1})\stackrel{(b)}=I(X_{r},X_{2};Y|X_{1},V_{1}),
\end{equation}
\begin{equation}\label{appen1.10.3}
R_{1}+R^{*}_{1}+R_{2}+R^{*}_{2}\leq I(V_{1},V_{2},X_{r},X_{1},X_{2};Y)\stackrel{(c)}=I(X_{r},X_{1},X_{2};Y),
\end{equation}
where (a) is from the Markov chain $V_{1}\rightarrow (X_{r},X_{1},X_{2},V_{2})\rightarrow Y$, (b)
is from the Markov chain $V_{2}\rightarrow (X_{r},X_{1},X_{2},V_{1})\rightarrow Y$, and (c) is from
the Markov chain $(V_{1},V_{2})\rightarrow (X_{r},X_{1},X_{2})\rightarrow Y$.

By using (\ref{appen1.7.1}), (\ref{appen1.7.2}), (\ref{appen1.7.3}), (\ref{appen1.10.1}), (\ref{appen1.10.2})
and (\ref{appen1.10.3}), it is easy to check that $P_{e}\leq\epsilon$. It remains to show that $\Delta\geq R_{1}+R_{2}-\epsilon$, see
the followings.

\textbf{Equivocation Analysis:}

Similar to the equivocation analysis of \cite[proof of Theorem 2]{LG}, for simplicity,
we only focus on the equivocation of one block, see the followings.

\begin{eqnarray}\label{appen1.14}
\frac{1}{N}H(W_{1},W_{2}|Z^{N})&=&\frac{1}{N}(H(W_{1},W_{2},Z^{N})-H(Z^{N}))\nonumber\\
&=&\frac{1}{N}(H(W_{1},W_{2},Z^{N},X_{1}^{N},X_{2}^{N})-H(X_{1}^{N},X_{2}^{N}|W_{1},W_{2},Z^{N})-H(Z^{N}))\nonumber\\
&\stackrel{(a)}=&\frac{1}{N}(H(Z^{N}|X_{1}^{N},X_{2}^{N})+H(X_{1}^{N})+H(X_{2}^{N})-H(X_{1}^{N},X_{2}^{N}|W_{1},W_{2},Z^{N})-H(Z^{N}))\nonumber\\
&=&\frac{1}{N}(H(X_{1}^{N})+H(X_{2}^{N})-I(X_{1}^{N},X_{2}^{N};Z^{N})-H(X_{1}^{N},X_{2}^{N}|W_{1},W_{2},Z^{N})),
\end{eqnarray}
where (a) follows from $(W_{1},W_{2})\rightarrow (X_{1}^{N},X_{2}^{N})\rightarrow Z^{N}$, $H(W_{1}|X_{1}^{N})=0$, $H(W_{2}|X_{2}^{N})=0$, and
$X_{1}^{N}$ is independent of $X_{2}^{N}$.

Consider the first term of (\ref{appen1.14}), the code-book generation of $x_{1}^{N}$ shows that the total number of $x_{1}^{N}$ is
$2^{N(R_{1}+R_{1}^{*})}$. Thus, using the same approach as that in \cite[Lemma 3]{LP}, we have
\begin{equation}\label{appen1.15}
\frac{1}{N}H(X_{1}^{N})\geq R_{1}+R_{1}^{*}-\epsilon_{1,N},
\end{equation}
where $\epsilon_{1,N}\rightarrow 0$ as $N\rightarrow \infty$.

Analogously, the second term of (\ref{appen1.14}) is bounded by
\begin{equation}\label{appen1.15.1}
\frac{1}{N}H(X_{2}^{N})\geq R_{2}+R_{2}^{*}-\epsilon_{2,N},
\end{equation}
where $\epsilon_{2,N}\rightarrow 0$ as $N\rightarrow \infty$.

For the third term of (\ref{appen1.14}), since the channel is memoryless, and $X_{1}^{N}$, $X_{2}^{N}$, $X_{r}^{N}$ are i.i.d. generated,
we get
\begin{equation}\label{appen1.16}
\frac{1}{N}I(X_{1}^{N},X_{2}^{N};Z^{N})=I(X_{1},X_{2};Z).
\end{equation}

Now, we consider the last term of (\ref{appen1.14}). Given $W_{1}$ and $W_{2}$, the wiretapper does joint decoding at each block.
At the end of block $1$, the wiretapper tries to find a unique pair $(\tilde{w}^{*}_{1,1}, \tilde{w}^{*}_{2,1})$ such that \\
$(x_{1}^{N}(w_{1,1},\tilde{w}^{*}_{1,1}|1,1),x_{2}^{N}(w_{2,1},\tilde{w}^{*}_{2,1}|1,1),z^{N}(1))$ are jointly typical.
At the end of block $i$ $(2\leq i\leq n-1)$, the wiretapper already has an estimation of the $\tilde{w}^{*}_{1,i-1}$ and $\tilde{w}^{*}_{2,i-1}$,
and thus he also get $\tilde{s}_{1,i}=(w_{1,i-1},\tilde{w}^{*}_{1,i-1})$ and $\tilde{s}_{2,i}=(w_{2,i-1},\tilde{w}^{*}_{2,i-1})$. Then he
tries to find a unique pair $(\tilde{w}^{*}_{1,i}, \tilde{w}^{*}_{2,i})$ such that
$(x_{1}^{N}(w_{1,i},\tilde{w}^{*}_{1,i}|\tilde{s}_{1,i}),x_{2}^{N}(w_{2,i},\tilde{w}^{*}_{2,i}|\tilde{s}_{2,i}),z^{N}(i))$ are jointly typical.
Based on the AEP, the error probability $Pr\{(\tilde{w}^{*}_{1,i},\tilde{w}^{*}_{2,i})\neq (w^{*}_{1,i},w^{*}_{2,i})\}$
goes to $0$ if
\begin{equation}\label{appen1.17.1}
R^{*}_{1}\leq I(X_{1};Z|X_{2}),
\end{equation}
\begin{equation}\label{appen1.17.2}
R^{*}_{2}\leq I(X_{2};Z|X_{1}),
\end{equation}
\begin{equation}\label{appen1.17.3}
R^{*}_{1}+R^{*}_{2}\leq I(X_{1},X_{2};Z).
\end{equation}
Then based on Fano¡¯s inequality, we have
\begin{equation}\label{appen1.17}
\frac{1}{N}H(X_{1}^{N},X_{2}^{N}|W_{1},W_{2},Z^{N})\leq \epsilon_{3,N},
\end{equation}
where $\epsilon_{3,N}\rightarrow 0$ as $N\rightarrow \infty$.

Substituting (\ref{appen1.15}), (\ref{appen1.15.1}), (\ref{appen1.16}) and (\ref{appen1.17}) into (\ref{appen1.14}), we have
\begin{equation}\label{appen1.18}
\frac{1}{N}H(W_{1},W_{2}|Z^{N})\geq R_{1}+R_{1}^{*}+R_{2}+R_{2}^{*}-I(X_{1},X_{2};Z)-\epsilon_{1,N}-\epsilon_{2,N}-\epsilon_{3,N}.
\end{equation}
It is easy to see that if we let
\begin{equation}\label{appen1.17.r}
R_{1}^{*}+R_{2}^{*}=I(X_{1},X_{2};Z),
\end{equation}
and choose sufficiently large $N$ such that $\epsilon_{1,N}+\epsilon_{2,N}+\epsilon_{3,N}\leq \epsilon$,
$\Delta=\frac{1}{N}H(W_{1},W_{2}|Z^{N})\geq R_{1}+R_{2}-\epsilon$ is guaranteed.

Based on (\ref{appen1.7.1}), (\ref{appen1.7.2}), (\ref{appen1.7.3}), (\ref{appen1.10.1}), (\ref{appen1.10.2}), (\ref{appen1.10.3}),
(\ref{appen1.17.1}), (\ref{appen1.17.2}) and (\ref{appen1.17.r}), the achievable region $\mathcal{R}^{d1}$ is obtained.

The proof of Theorem \ref{T1} is completed.

\section{Proof of Theorem \ref{T2}\label{appen2}}

For Theorem \ref{T2}, we only need to prove that the corner points of $\mathcal{L}^{1}$
and $\mathcal{L}^{2}$ are achievable, see the followings.

\begin{itemize}

\item \textbf{(Case 1)} If $I(X_{r};Y)\geq I(X_{r};Z)$,
we allow the legitimate receiver to decode $x_{r}^{N}$, and the wiretapper can not decode it.
For case 1, it is sufficient to show that the pair $(R_{1},R_{2})\in \mathcal{L}^{1}$ with the condition
\begin{equation}\label{appen5.1}
R_{1}=I(X_{1};Y|X_{2},X_{r})-I(X_{1},X_{r};Z)+R_{r}, \,\,\, R_{2}=I(X_{2};Y|X_{r})-I(X_{2};Z|X_{1},X_{r})
\end{equation}
is achievable. The achievability proof of the other corner point $(R_{1}=I(X_{1};Y|X_{r})-I(X_{1};Z|X_{2},X_{r}), R_{2}=I(X_{2};Y|X_{1},X_{r})-I(X_{2},X_{r};Z)+R_{r})$
follows by symmetry.

\item \textbf{(Case 2)} If $I(X_{r};Y)\leq I(X_{r};Z)$,
we allow both the receivers to decode $x_{r}^{N}$.
For case 2, it is sufficient to show that the pair $(R_{1},R_{2})\in \mathcal{L}^{2}$ with the condition
\begin{equation}\label{appen5.2}
R_{1}=I(X_{1};Y|X_{2},X_{r})-I(X_{1};Z|X_{r}), \,\,\, R_{2}=I(X_{2};Y|X_{r})-I(X_{2};Z|X_{1},X_{r})
\end{equation}
is achievable. The achievability proof of the other corner point $(R_{1}=I(X_{1};Y|X_{r})-I(X_{1};Z|X_{2},X_{r}), R_{2}=I(X_{2};Y|X_{1},X_{r})-I(X_{2};Z|X_{r}))$
follows by symmetry.

\end{itemize}

Fix the joint probability mass function
$P_{Y,Z,Y_{r}|X_{r},X_{1},X_{2}}(y,z,y_{r}|x_{r},x_{1},x_{2})P_{X_{r}}(x_{r})P_{X_{1}}(x_{1})P_{X_{2}}(x_{2})$.
Define the messages $W_{1}$, $W_{2}$ taking values in the alphabets
$\mathcal{W}_{1}$, $\mathcal{W}_{2}$, respectively, where
$$\mathcal{W}_{1}=\{1,2,...,2^{NR_{1}}\}, \,\,\, \mathcal{W}_{2}=\{1,2,...,2^{NR_{2}}\}.$$

\textbf{Code-book Construction for the Two Cases:}

\begin{itemize}

\item \textbf{Code-book construction for case 1}:

\begin{itemize}

\item First, generate at random $2^{N(R_{r}-\epsilon^{'})}$ (where $\epsilon^{'}$ is a small positive number)
i.i.d. sequences at the relay node each drawn according to
$P_{X_{r}^{N}}(x_{r}^{N})=\prod_{i=1}^{N}P_{X_{r}}(x_{r,i})$, index them as $x_{r}^{N}(a)$, $a\in [1,2^{N(R_{r}-\epsilon^{'})}]$, where
\begin{equation}\label{appen5.11}
R_{r}=\min\{I(X_{r};Z|X_{1}), I(X_{r};Z|X_{2}), I(X_{r};Y)\}.
\end{equation}
Here note that
\begin{equation}\label{appen5.12}
R_{r}\geq I(X_{r};Z).
\end{equation}

\item Second, generate $2^{N(I(X_{2};Y|X_{r})-\epsilon^{'})}$ i.i.d. codewords $x_{2}^{N}$ according to $P_{X_{2}}(x_{2})$,
and divide them into $2^{NR_{2}}$ bins. Each bin
contains $2^{N(I(X_{2};Y|X_{r})-\epsilon^{'}-R_{2})}$ codewords,
where
\begin{equation}\label{appen5.13}
I(X_{2};Y|X_{r})-\epsilon^{'}-R_{2}=I(X_{2};Z|X_{1},X_{r})-\epsilon^{'}.
\end{equation}

\item Third, generate $2^{N(I(X_{1};Y|X_{2},X_{r})-\epsilon^{'})}$ i.i.d. codewords $x_{1}^{N}$ according to $P_{X_{1}}(x_{1})$,
and divide them into $2^{NR_{1}}$ bins. Each bin
contains $2^{N(I(X_{1};Y|X_{2},X_{r})-\epsilon^{'}-R_{1})}$ codewords.

\end{itemize}

\item \textbf{Code-book Construction for case 2}:

\begin{itemize}

\item Generate at random $2^{N(R_{r}-\epsilon^{'})}$
i.i.d. sequences at the relay node each drawn according to
$P_{X_{r}^{N}}(x_{r}^{N})=\prod_{i=1}^{N}P_{X_{r}}(x_{r,i})$, index them as $x_{r}^{N}(a)$, $a\in [1,2^{N(R_{r}-\epsilon^{'})}]$, where
\begin{equation}\label{appen5.14}
R_{r}=I(X_{r};Y)\leq I(X_{r};Z).
\end{equation}

\item Second, generate $2^{N(I(X_{2};Y|X_{r})-\epsilon^{'})}$ i.i.d. codewords $x_{2}^{N}$ according to $P_{X_{2}}(x_{2})$,
and divide them into $2^{NR_{2}}$ bins. Each bin
contains $2^{N(I(X_{2};Y|X_{r})-\epsilon^{'}-R_{2})}$ codewords, where
\begin{equation}\label{appen5.15}
I(X_{2};Y|X_{r})-\epsilon^{'}-R_{2}=I(X_{2};Z|X_{1},X_{r})-\epsilon^{'}.
\end{equation}

\item Third, generate $2^{N(I(X_{1};Y|X_{2},X_{r})-\epsilon^{'})}$ i.i.d. codewords $x_{1}^{N}$ according to $P_{X_{1}}(x_{1})$,
and divide them into $2^{NR_{1}}$ bins. Each bin
contains $2^{N(I(X_{1};Y|X_{2},X_{r})-\epsilon^{'}-R_{1})}$ codewords, where
\begin{equation}\label{appen5.16}
I(X_{1};Y|X_{2},X_{r})-\epsilon^{'}-R_{1}=I(X_{1};Z|X_{r})-\epsilon^{'}.
\end{equation}

\end{itemize}

\end{itemize}

\textbf{Encoding for both cases:}

The relay uniformly picks a codeword $x_{r}^{N}(a)$ from $[1,2^{N(R_{r}-\epsilon^{'})}]$, and sends $x_{r}^{N}(a)$.

For a given confidential message $w_{2}$, randomly choose a
codeword $x_{2}^{N}$ in bin $w_{2}$ to transmit. Similarly, for a given
confidential message $w_{1}$, randomly choose a codeword $x_{1}^{N}$
in bin $w_{1}$ to transmit.

\textbf{Decoding for both cases:}

For a given $y^{N}$, try to find a sequence $x_{r}^{N}(\hat{a})$ such that
$(x_{r}^{N}(\hat{a}),y^{N})$ are jointly typical. If there exists a unique
sequence with the index $\hat{a}$, put out the corresponding $\hat{a}$, else
declare a decoding error. Based on the AEP and (\ref{appen5.11}) (or (\ref{appen5.14})),
the probability $Pr\{\hat{a}=a\}$ goes to 1.

After decoding $\hat{a}$, the legitimate receiver tries to find
a sequence $x_{2}^{N}(\hat{w}_{2})$ such that $(x_{2}^{N}(\hat{w}_{2}),x_{r}^{N}(\hat{a}),y^{N})$ are jointly typical.
If there exists a unique sequence with the index
$\hat{w}_{2}$, put out the corresponding $\hat{w}_{2}$, else declare a decoding
error. Based on the AEP and the construction of $x_{2}^{N}$ for both cases, the
probability $Pr\{\hat{w}_{2}=w_{2}\}$ goes to 1.

Finally, after decoding $\hat{a}$ and $\hat{w}_{2}$, the legitimate
receiver tries to find a sequence $x_{1}^{N}(\hat{w}_{1})$ such that \\
$(x_{1}^{N}(\hat{w}_{1}),x_{2}^{N}(\hat{w}_{2}),x_{r}^{N}(\hat{a}),y^{N})$ are jointly typical.
If there
exists a unique sequence with the index $\hat{w}_{1}$, put out the
corresponding $\hat{w}_{1}$, else declare a decoding error. Based
on the AEP and the construction of $x_{1}^{N}$ for both cases, the probability
$Pr\{\hat{w}_{1}=w_{1}\}$ goes to 1.

$P_{e}\leq\epsilon$ is easy to be checked by using the above encoding-decoding schemes.
Now, it remains to prove $\Delta\geq R_{1}+R_{2}-\epsilon$ for both cases, see the followings.

\textbf{Equivocation Analysis:}

\textbf{Proof of $\Delta\geq R_{1}+R_{2}-\epsilon$ for case 1:}

\begin{eqnarray}\label{appen5.17}
\Delta&=&\frac{1}{N}H(W_{1},W_{2}|Z^{N})\nonumber\\
&=&\frac{1}{N}(H(W_{1}|Z^{N})+H(W_{2}|W_{1},Z^{N})).
\end{eqnarray}
The first term in (\ref{appen5.17}) is bounded as follows.

\begin{eqnarray}\label{appen5.18}
\frac{1}{N}H(W_{1}|Z^{N})&=&\frac{1}{N}(H(W_{1},Z^{N})-H(Z^{N}))\nonumber\\
&=&\frac{1}{N}(H(W_{1},Z^{N},X_{1}^{N},X_{r}^{N})-H(X_{1}^{N},X_{r}^{N}|W_{1},Z^{N})-H(Z^{N}))\nonumber\\
&\stackrel{(a)}=&\frac{1}{N}(H(Z^{N}|X_{1}^{N},X_{r}^{N})+H(X_{1}^{N})+H(X_{r}^{N})-H(X_{1}^{N},X_{r}^{N}|W_{1},Z^{N})-H(Z^{N}))\nonumber\\
&=&\frac{1}{N}(H(X_{1}^{N})+H(X_{r}^{N})-I(X_{1}^{N},X_{r}^{N};Z^{N})-H(X_{1}^{N},X_{r}^{N}|W_{1},Z^{N})),
\end{eqnarray}
where (a) follows from $W_{1}\rightarrow (X_{1}^{N},X_{r}^{N})\rightarrow Z^{N}$, $H(W_{1}|X_{1}^{N})=0$ and the fact that $X_{1}^{N}$
is independent of $X_{r}^{N}$.

Consider the first term in (\ref{appen5.18}), the code-book generation of $x_{1}^{N}$ shows that the total number of $x_{1}^{N}$ is
$2^{N(I(X_{1};Y|X_{2},X_{r})-\epsilon^{'})}$. Thus,
using the same approach as that in \cite[Lemma 3]{LP}, we have
\begin{equation}\label{appen5.19}
\frac{1}{N}H(X_{1}^{N})\geq I(X_{1};Y|X_{2},X_{r})-\epsilon^{'}-\epsilon_{1,N},
\end{equation}
where $\epsilon_{1,N}\rightarrow 0$ as $N\rightarrow \infty$.

For the second term in (\ref{appen5.18}), the code-book generation of $x_{r}^{N}$ guarantees that
\begin{equation}\label{appen5.20}
\frac{1}{N}H(X_{r}^{N})\geq R_{r}-\epsilon^{'}-\epsilon_{2,N},
\end{equation}
where $\epsilon_{2,N}\rightarrow 0$ as $N\rightarrow \infty$.

For the third term in (\ref{appen5.18}), since the channel is memoryless, and $X_{1}^{N}$, $X_{2}^{N}$, $X_{r}^{N}$ are i.i.d. generated, we get
\begin{equation}\label{appen5.21}
\frac{1}{N}I(X_{1}^{N},X_{r}^{N};Z^{N})=I(X_{1},X_{r};Z).
\end{equation}

Now, we consider the last term of (\ref{appen5.18}). Given $w_{1}$, the
wiretapper can do joint decoding. Specifically, given
$z^{N}$ and $w_{1}$,
\begin{equation}\label{appen5.22}
\frac{1}{N}H(X_{1}^{N},X_{r}^{N}|W_{1},Z^{N})\leq \epsilon_{3,N}
\end{equation}
($\epsilon_{3,N}\rightarrow 0$ as $N\rightarrow \infty$) is guaranteed if $R_{r}\leq I(X_{r};Z|X_{1})$ and $R_{r}\geq I(X_{r};Z)$, and
this is from the properties of AEP (similar argument is used
in the proof of \cite[Theorem 3]{LG}). By checking (\ref{appen5.11}) and (\ref{appen5.12}),
(\ref{appen5.22}) is obtained.

Substituting (\ref{appen5.19}), (\ref{appen5.20}), (\ref{appen5.21}) and (\ref{appen5.22}) into (\ref{appen5.18}), we have
\begin{equation}\label{appen5.23}
\frac{1}{N}H(W_{1}|Z^{N})\geq I(X_{1};Y|X_{2},X_{r})+R_{r}-I(X_{1},X_{r};Z)-2\epsilon^{'}-\epsilon_{1,N}-\epsilon_{2,N}-\epsilon_{3,N}.
\end{equation}

The second term in (\ref{appen5.17}) is bounded as follows.

\begin{eqnarray}\label{appen5.24}
\frac{1}{N}H(W_{2}|W_{1},Z^{N})&\geq&\frac{1}{N}H(W_{2}|W_{1},Z^{N},X_{1}^{N},X_{r}^{N})\nonumber\\
&\stackrel{(1)}=&\frac{1}{N}H(W_{2}|Z^{N},X_{1}^{N},X_{r}^{N})\nonumber\\
&=&\frac{1}{N}(H(W_{2},Z^{N},X_{1}^{N},X_{r}^{N})-H(Z^{N},X_{1}^{N},X_{r}^{N}))\nonumber\\
&=&\frac{1}{N}(H(W_{2},Z^{N},X_{1}^{N},X_{r}^{N},X_{2}^{N})-H(X_{2}^{N}|W_{2},Z^{N},X_{1}^{N},X_{r}^{N})-H(Z^{N},X_{1}^{N},X_{r}^{N}))\nonumber\\
&\stackrel{(2)}=&\frac{1}{N}(H(Z^{N}|X_{1}^{N},X_{2}^{N},X_{r}^{N})+H(X_{r}^{N})+H(X_{1}^{N})+H(X_{2}^{N})\nonumber\\
&&-H(X_{2}^{N}|W_{2},Z^{N},X_{1}^{N},X_{r}^{N})-H(Z^{N}|X_{1}^{N},X_{r}^{N})-H(X_{1}^{N})-H(X_{r}^{N}))\nonumber\\
&=&\frac{1}{N}(H(X_{2}^{N})-I(X_{2}^{N};Z^{N}|X_{1}^{N},X_{r}^{N})-H(X_{2}^{N}|W_{2},Z^{N},X_{1}^{N},X_{r}^{N})),
\end{eqnarray}
where (1) is from the Markov chain $W_{1}\rightarrow (Z^{N},X_{1}^{N},X_{r}^{N})\rightarrow W_{2}$, and (2) is from
the Markov chain $W_{2}\rightarrow (X_{1}^{N},X_{2}^{N},X_{r}^{N})\rightarrow Z^{N}$, $H(W_{2}|X_{2}^{N})=0$, and the fact that
$X_{1}^{N}$, $X_{2}^{N}$ and $X_{r}^{N}$ are independent.

Consider the first term in (\ref{appen5.24}), the code-book generation of $x_{2}^{N}$ shows that the total number of $x_{2}^{N}$ is
$2^{N(I(X_{2};Y|X_{r})-\epsilon^{'})}$. Thus,
using the same approach as that in \cite[Lemma 3]{LP}, we have
\begin{equation}\label{appen5.25}
\frac{1}{N}H(X_{2}^{N})\geq I(X_{2};Y|X_{r})-\epsilon^{'}-\epsilon_{4,N},
\end{equation}
where $\epsilon_{4,N}\rightarrow 0$ as $N\rightarrow \infty$.

For the second term in (\ref{appen5.24}), since the channel is memoryless, and $X_{1}^{N}$, $X_{2}^{N}$, $X_{r}^{N}$ are i.i.d. generated, we get
\begin{equation}\label{appen5.26}
\frac{1}{N}I(X_{2}^{N};Z^{N}|X_{1}^{N},X_{r}^{N})=I(X_{2};Z|X_{1},X_{r}).
\end{equation}

Now, we consider the last term of (\ref{appen5.24}). Given $Z^{N}$, $X_{1}^{N}$, $X_{r}^{N}$ and $W_{2}$,
the total number of possible codewords of $x_{2}^{N}$ is $2^{N(I(X_{2};Y|X_{r})-\epsilon^{'}-R_{2})}$.
By using the Fano's inequality and (\ref{appen5.13}),
we have
\begin{equation}\label{appen5.27}
\frac{1}{N}H(X_{2}^{N}|W_{2},Z^{N},X_{1}^{N},X_{r}^{N})\leq \epsilon_{5,N},
\end{equation}
where $\epsilon_{5,N}\rightarrow 0$ as $N\rightarrow \infty$.

Substituting (\ref{appen5.25}), (\ref{appen5.26}) and (\ref{appen5.27}) into (\ref{appen5.24}), we have
\begin{equation}\label{appen5.28}
\frac{1}{N}H(W_{2}|W_{1},Z^{N})\geq I(X_{2};Y|X_{r})-I(X_{2};Z|X_{1},X_{r})-\epsilon^{'}-\epsilon_{4,N}-\epsilon_{5,N}.
\end{equation}

Substituting (\ref{appen5.23}) and (\ref{appen5.28}) into (\ref{appen5.17}),
and choosing $\epsilon^{'}$ and sufficiently large $N$ such that $3\epsilon^{'}+\epsilon_{1,N}+\epsilon_{2,N}+\epsilon_{3,N}+\epsilon_{4,N}+\epsilon_{5,N}\leq \epsilon$,
$\Delta\geq R_{1}+R_{2}-\epsilon$ for case 1 is proved.

\textbf{Proof of $\Delta\geq R_{1}+R_{2}-\epsilon$ for case 2:}

\begin{eqnarray}\label{appen5.29}
\Delta&=&\frac{1}{N}H(W_{1},W_{2}|Z^{N})\nonumber\\
&=&\frac{1}{N}(H(W_{1}|Z^{N})+H(W_{2}|W_{1},Z^{N})).
\end{eqnarray}
The first term in (\ref{appen5.29}) is bounded as follows.

\begin{eqnarray}\label{appen5.30}
\frac{1}{N}H(W_{1}|Z^{N})&\geq&\frac{1}{N}H(W_{1}|Z^{N},X_{r}^{N})\nonumber\\
&=&\frac{1}{N}(H(W_{1},Z^{N},X_{r}^{N})-H(Z^{N},X_{r}^{N}))\nonumber\\
&=&\frac{1}{N}(H(W_{1},Z^{N},X_{1}^{N},X_{r}^{N})-H(X_{1}^{N}|W_{1},Z^{N},X_{r}^{N})-H(Z^{N},X_{r}^{N}))\nonumber\\
&\stackrel{(a)}=&\frac{1}{N}(H(Z^{N}|X_{1}^{N},X_{r}^{N})+H(X_{1}^{N})+H(X_{r}^{N})-H(X_{1}^{N}|W_{1},Z^{N},X_{r}^{N})\nonumber\\
&&-H(Z^{N}|X_{r}^{N})-H(X_{r}^{N}))\nonumber\\
&=&\frac{1}{N}(H(X_{1}^{N})-I(X_{1}^{N};Z^{N}|X_{r}^{N})-H(X_{1}^{N}|W_{1},Z^{N},X_{r}^{N})),
\end{eqnarray}
where (a) follows from $W_{1}\rightarrow (X_{1}^{N},X_{r}^{N})\rightarrow Z^{N}$, $H(W_{1}|X_{1}^{N})=0$ and the fact that $X_{1}^{N}$
is independent of $X_{r}^{N}$.

Consider the first term in (\ref{appen5.30}), the code-book generation of $x_{1}^{N}$ shows that the total number of $x_{1}^{N}$ is
$2^{N(I(X_{1};Y|X_{2},X_{r})-\epsilon^{'})}$. Thus,
using the same approach as that in \cite[Lemma 3]{LP}, we have
\begin{equation}\label{appen5.31}
\frac{1}{N}H(X_{1}^{N})\geq I(X_{1};Y|X_{2},X_{r})-\epsilon^{'}-\epsilon_{1,N},
\end{equation}
where $\epsilon_{1,N}\rightarrow 0$ as $N\rightarrow \infty$.

For the second term in (\ref{appen5.30}), since the channel is memoryless, and $X_{1}^{N}$, $X_{2}^{N}$, $X_{r}^{N}$ are i.i.d. generated, we get
\begin{equation}\label{appen5.32}
\frac{1}{N}I(X_{1}^{N};Z^{N}|X_{r}^{N})=I(X_{1};Z|X_{r}).
\end{equation}

Now, we consider the last term of (\ref{appen5.30}). Given $Z^{N}$, $X_{r}^{N}$ and $W_{1}$,
the total number of possible codewords of $x_{1}^{N}$ is $2^{N(I(X_{1};Y|X_{2},X_{r})-\epsilon^{'}-R_{1})}$.
By using the Fano's inequality and (\ref{appen5.16}),
we have
\begin{equation}\label{appen5.33}
\frac{1}{N}H(X_{1}^{N}|W_{1},Z^{N},X_{r}^{N})\leq \epsilon_{2,N},
\end{equation}
where $\epsilon_{2,N}\rightarrow 0$ as $N\rightarrow \infty$.

Substituting (\ref{appen5.31}), (\ref{appen5.32}) and (\ref{appen5.33}) into (\ref{appen5.30}), we have
\begin{equation}\label{appen5.34}
\frac{1}{N}H(W_{1}|Z^{N})\geq I(X_{1};Y|X_{2},X_{r})-I(X_{1};Z|X_{r})-\epsilon^{'}-\epsilon_{1,N}-\epsilon_{2,N}.
\end{equation}

The second term in (\ref{appen5.29}) is bounded the same as that for case 1, and thus, we have
\begin{equation}\label{appen5.35}
\frac{1}{N}H(W_{2}|W_{1},Z^{N})\geq I(X_{2};Y|X_{r})-I(X_{2};Z|X_{1},X_{r})-\epsilon^{'}-\epsilon_{3,N}-\epsilon_{4,N},
\end{equation}
where $\epsilon_{3,N}, \epsilon_{4,N}\rightarrow 0$ as $N\rightarrow \infty$.
The proof is omitted here.

Substituting (\ref{appen5.34}) and (\ref{appen5.35}) into (\ref{appen5.29}),
and choosing $\epsilon^{'}$ and sufficiently large $N$ such that $2\epsilon^{'}+\epsilon_{1,N}+\epsilon_{2,N}+\epsilon_{3,N}+\epsilon_{4,N}\leq \epsilon$,
$\Delta\geq R_{1}+R_{2}-\epsilon$ for case 2 is proved.

The proof of Theorem \ref{T2} is completed.

\section{Proof of Theorem \ref{T3}\label{appen3}}

For Theorem \ref{T3}, we only need to prove that the corner points of $\mathcal{L}^{3}$
and $\mathcal{L}^{4}$ are achievable, see the followings.

\begin{itemize}

\item \textbf{(Case 1)} If $I(X_{r};Y)\geq I(X_{r};Z)$,
we allow the legitimate receiver to decode $x_{r}^{N}$, and the wiretapper can not decode it.
For case 1, it is sufficient to show that the pair $(R_{1},R_{2})\in \mathcal{L}^{3}$ with the condition
\begin{equation}\label{appen6.1}
R_{1}=I(X_{1};Y,\hat{Y}_{r}|X_{2},X_{r})-I(X_{1},X_{r};Z)+R^{*}, \,\,\, R_{2}=I(X_{2};Y,\hat{Y}_{r}|X_{r})-I(X_{2};Z|X_{1},X_{r})
\end{equation}
is achievable. The achievability proof of the other corner point $(R_{1}=I(X_{1};Y,\hat{Y}_{r}|X_{r})-I(X_{1};Z|X_{2},X_{r}),
R_{2}=I(X_{2};Y,\hat{Y}_{r}|X_{1},X_{r})-I(X_{2},X_{r};Z)+R^{*})$
follows by symmetry. Here note that $R^{*}$ satisfies
\begin{equation}\label{appen6.2}
\min\{I(X_{r};Z|X_{1}), I(X_{r};Z|X_{2}), I(X_{r};Y)\}-R^{*}\geq I(Y_{r};\hat{Y}_{r}|X_{r}).
\end{equation}

\item \textbf{(Case 2)} If $I(Y_{r};\hat{Y}_{r}|X_{r})\leq I(X_{r};Y)\leq I(X_{r};Z)$,
we allow both the receivers to decode $x_{r}^{N}$.
For case 2, it is sufficient to show that the pair $(R_{1},R_{2})\in \mathcal{L}^{4}$ with the condition
\begin{equation}\label{appen6.4}
R_{1}=I(X_{1};Y,\hat{Y}_{r}|X_{2},X_{r})-I(X_{1};Z|X_{r}), \,\,\, R_{2}=I(X_{2};Y,\hat{Y}_{r}|X_{r})-I(X_{2};Z|X_{1},X_{r})
\end{equation}
is achievable. The achievability proof of the other corner point $(R_{1}=I(X_{1};Y,\hat{Y}_{r}|X_{r})-I(X_{1};Z|X_{2},X_{r}), R_{2}=I(X_{2};Y,\hat{Y}_{r}|X_{1},X_{r})-I(X_{2};Z|X_{r}))$
follows by symmetry.

\end{itemize}

Fix the joint probability mass function
$P_{\hat{Y}_{r}|Y_{r},X_{r}}(\hat{y}_{r}|y_{r},x_{r})P_{Y,Z,Y_{r}|X_{r},X_{1},X_{2}}(y,z,y_{r}|x_{r},x_{1},x_{2})
P_{X_{r}}(x_{r})P_{X_{1}}(x_{1})P_{X_{2}}(x_{2})$.
Define the messages $W_{1}$, $W_{2}$ taking values in the alphabets
$\mathcal{W}_{1}$, $\mathcal{W}_{2}$, respectively, where
$$\mathcal{W}_{1}=\{1,2,...,2^{NR_{1}}\}, \,\,\, \mathcal{W}_{2}=\{1,2,...,2^{NR_{2}}\}.$$

\textbf{Code-book Construction for the Two Cases:}

\begin{itemize}

\item \textbf{Code-book construction for case 1}:

\begin{itemize}

\item First, generate at random $2^{N(R^{*}_{r1}-\epsilon^{'})}$ ($\epsilon^{'}$ is a small positive number)
i.i.d. sequences $x_{r}^{N}$ at the relay node each drawn according to
$P_{X_{r}^{N}}(x_{r}^{N})=\prod_{i=1}^{N}P_{X_{r}}(x_{r,i})$, index them as $x_{r}^{N}(a)$, $a\in [1,2^{N(R^{*}_{r1}-\epsilon^{'})}]$, where
\begin{equation}\label{appen6.5}
R^{*}_{r1}=\min\{I(X_{r};Z|X_{1}), I(X_{r};Z|X_{2}), I(X_{r};Y)\}.
\end{equation}
Here note that
\begin{equation}\label{appen6.6}
R^{*}_{r1}\geq I(X_{r};Z).
\end{equation}
For each $x_{r}^{N}(a)$ ($a\in [1,2^{N(R^{*}_{r1}-\epsilon^{'})}]$), generate at random $2^{N(R^{*}_{r1}-\epsilon^{'}-R^{*})}$ i.i.d.
$\hat{y}_{r}^{N}$ according to $P_{\hat{Y}_{r}^{N}|X^{N}_{r}}(\hat{y}_{r}^{N}|x^{N}_{r})=\prod_{i=1}^{N}P_{\hat{Y}_{r}|X_{r}}(\hat{y}_{r,i}|x_{r,i})$.
Label these $\hat{y}_{r}^{N}$ as $\hat{y}_{r}^{N}(m,a)$, $m\in [1,2^{N(R^{*}_{r1}-\epsilon^{'}-R^{*})}]$, $a\in [1,2^{N(R^{*}_{r1}-\epsilon^{'})}]$.
Equally divide $2^{N(R^{*}_{r1}-\epsilon^{'})}$ sequences of $x_{r}^{N}$ into $2^{N(R^{*}_{r1}-\epsilon^{'}-R^{*})}$ bins, hence there are
$2^{NR^{*}}$ sequences of $x_{r}^{N}$ at each bin.

\item Second, generate $2^{N(I(X_{2};Y,\hat{Y}_{r}|X_{r})-\epsilon^{'})}$ i.i.d. codewords $x_{2}^{N}$ according to $P_{X_{2}}(x_{2})$,
and divide them into $2^{NR_{2}}$ bins. Each bin
contains $2^{N(I(X_{2};Y,\hat{Y}_{r}|X_{r})-\epsilon^{'}-R_{2})}$ codewords,
where
\begin{equation}\label{appen6.7}
I(X_{2};Y,\hat{Y}_{r}|X_{r})-\epsilon^{'}-R_{2}=I(X_{2};Z|X_{1},X_{r})-\epsilon^{'}.
\end{equation}

\item Third, generate $2^{N(I(X_{1};Y,\hat{Y}_{r}|X_{2},X_{r})-\epsilon^{'}+R^{*}-R^{*}_{r1})}$ i.i.d. codewords $x_{1}^{N}$ according to $P_{X_{1}}(x_{1})$,
and divide them into $2^{NR_{1}}$ bins. Each bin
contains $2^{N(I(X_{1};Y,\hat{Y}_{r}|X_{2},X_{r})-\epsilon^{'}+R^{*}-R^{*}_{r1}-R_{1})}$ codewords.
Here note that from (\ref{appen6.2}) and (\ref{appen6.5}), we know that $R^{*}\leq R^{*}_{r1}$, and thus, we have
\begin{equation}\label{appen6.71}
I(X_{1};Y,\hat{Y}_{r}|X_{2},X_{r})-\epsilon^{'}+R^{*}-R^{*}_{r1}\leq I(X_{1};Y,\hat{Y}_{r}|X_{2},X_{r})-\epsilon^{'}.
\end{equation}
In addition, by using $R_{1}=I(X_{1};Y,\hat{Y}_{r}|X_{2},X_{r})-I(X_{1},X_{r};Z)+R^{*}$,
the codewords $x_{1}^{N}$ in each bin is upper bounded by
\begin{eqnarray}\label{appen6.72}
&&I(X_{1};Y,\hat{Y}_{r}|X_{2},X_{r})-\epsilon^{'}+R^{*}-R^{*}_{r1}-R_{1}\nonumber\\
&=&I(X_{1};Y,\hat{Y}_{r}|X_{2},X_{r})-\epsilon^{'}+R^{*}-R^{*}_{r1}\nonumber\\
&&-(I(X_{1};Y,\hat{Y}_{r}|X_{2},X_{r})-I(X_{1},X_{r};Z)+R^{*})\nonumber\\
&=&I(X_{1},X_{r};Z)-R^{*}_{r1}-\epsilon^{'}\nonumber\\
&\stackrel{(a)}\leq&I(X_{1},X_{r};Z)-I(X_{r};Z)-\epsilon^{'}\nonumber\\
&=&I(X_{1};Z|X_{r})-\epsilon^{'},
\end{eqnarray}
where (a) is from (\ref{appen6.6}).

\end{itemize}

\item \textbf{Code-book Construction for case 2}:

\begin{itemize}

\item First, generate at random $2^{N(R^{*}_{r2}-\epsilon^{'})}$
i.i.d. sequences $x_{r}^{N}$ at the relay node each drawn according to
$P_{X_{r}^{N}}(x_{r}^{N})=\prod_{i=1}^{N}P_{X_{r}}(x_{r,i})$, index them as $x_{r}^{N}(a)$, $a\in [1,2^{N(R^{*}_{r2}-\epsilon^{'})}]$, where
\begin{equation}\label{appen6.8}
R^{*}_{r2}=I(X_{r};Y)\leq I(X_{r};Z).
\end{equation}

For each $x_{r}^{N}(a)$ ($a\in [1,2^{N(R^{*}_{r2}-\epsilon^{'})}]$), generate at random $2^{N(R^{*}_{r2}-\epsilon^{'})}$ i.i.d.
$\hat{y}_{r}^{N}$ according to $P_{\hat{Y}_{r}^{N}|X^{N}_{r}}(\hat{y}_{r}^{N}|x^{N}_{r})=\prod_{i=1}^{N}P_{\hat{Y}_{r}|X_{r}}(\hat{y}_{r,i}|x_{r,i})$.
Label these $\hat{y}_{r}^{N}$ as $\hat{y}_{r}^{N}(a)$, $a\in [1,2^{N(R^{*}_{r2}-\epsilon^{'})}]$.

\item Second, generate $2^{N(I(X_{2};Y,\hat{Y}_{r}|X_{r})-\epsilon^{'})}$ i.i.d. codewords $x_{2}^{N}$ according to $P_{X_{2}}(x_{2})$,
and divide them into $2^{NR_{2}}$ bins. Each bin
contains $2^{N(I(X_{2};Y,\hat{Y}_{r}|X_{r})-\epsilon^{'}-R_{2})}$ codewords, where
\begin{equation}\label{appen6.9}
I(X_{2};Y,\hat{Y}_{r}|X_{r})-\epsilon^{'}-R_{2}=I(X_{2};Z|X_{1},X_{r})-\epsilon^{'}.
\end{equation}

\item Third, generate $2^{N(I(X_{1};Y,\hat{Y}_{r}|X_{2},X_{r})-\epsilon^{'})}$ i.i.d. codewords $x_{1}^{N}$ according to $P_{X_{1}}(x_{1})$,
and divide them into $2^{NR_{1}}$ bins. Each bin
contains $2^{N(I(X_{1};Y,\hat{Y}_{r}|X_{2},X_{r})-\epsilon^{'}-R_{1})}$ codewords, where
\begin{equation}\label{appen6.10}
I(X_{1};Y,\hat{Y}_{r}|X_{2},X_{r})-\epsilon^{'}-R_{1}=I(X_{1};Z|X_{r})-\epsilon^{'}.
\end{equation}

\end{itemize}

\end{itemize}

\textbf{Encoding:}

Encoding involves the mapping of message indices to channel inputs, which are facilitated by the
sequences generated above. We exploit the block Markov coding scheme, as argued in \cite{CG}, the loss induced by
this scheme is negligible as the number of blocks $n\rightarrow \infty$. For block $i$ ($1\leq i\leq n$), encoding proceeds as follows.

First, for convenience, the messages $w_{1}$ and $w_{2}$ transmitted in the $i$-th block
are denoted by $w_{1,i}$ and $w_{2,i}$, respectively. $y_{r}^{N}(i)$ and $\hat{y}_{r}^{N}(i)$ are the $y_{r}^{N}$ and $\hat{y}_{r}^{N}$
for the $i$-th block, respectively.

\begin{itemize}

\item \textbf{Encoding for case 1:}

At the end of block $i$ ($2\leq i\leq n$), assume that $(x_{r}^{N}(a_{i}),y_{r}^{N}(i),\hat{y}_{r}^{N}(m_{i},a_{i}))$
are jointly typical, then we choose $a_{i+1}$ uniformly from bin $m_{i}$, and the relay sends $x_{r}^{N}(a_{i+1})$
at block $i+1$. In the first block, the relay sends $x_{r}^{N}(1)$.

For a given confidential message $w_{2}$, randomly choose a
codeword $x_{2}^{N}$ in bin $w_{2}$ to transmit. Similarly, for a given
confidential message $w_{1}$, randomly choose a codeword $x_{1}^{N}$
in bin $w_{1}$ to transmit.

\item \textbf{Encoding for case 2:}

In block $i$ ($1\leq i\leq n$), the relay randomly choose an index $a_{i}$ from $[1,2^{N(R^{*}_{r2}-\epsilon^{'})}]$, and sends
$x_{r}^{N}(a_{i})$ and $\hat{y}_{r}^{N}(a_{i})$.

For a given confidential message $w_{2}$, randomly choose a
codeword $x_{2}^{N}$ in bin $w_{2}$ to transmit. Similarly, for a given
confidential message $w_{1}$, randomly choose a codeword $x_{1}^{N}$
in bin $w_{1}$ to transmit.

\end{itemize}

\textbf{Decoding:}

\begin{itemize}

\item \textbf{Decoding for case 1:}

(At the relay) At the end of block $i$, the relay already has $a_{i}$, it then decides $m_{i}$ by choosing $m_{i}$ such that
$(x_{r}^{N}(a_{i}),y_{r}^{N}(i),\hat{y}_{r}^{N}(m_{i},a_{i}))$ are jointly typical. There exists such $m_{i}$, if
\begin{equation}\label{appen6.11}
R^{*}_{r1}-R^{*}\geq I(Y_{r};\hat{Y}_{r}|X_{r}),
\end{equation}
and $N$ is sufficiently large. Choose $a_{i+1}$ uniformly from bin $m_{i}$.

(At the legitimate receiver) The legitimate receiver does backward decoding. The decoding process starts at the last block $n$,
the legitimate receiver decodes $a_{n}$ by choosing unique $\check{a}_{n}$ such that $(x_{r}^{N}(\check{a}_{n}),y^{N}(n))$ are jointly typical. Since
$R^{*}_{r1}$ satisfies (\ref{appen6.5}), the probability $Pr\{\check{a}_{n}=a_{n}\}$ goes to $1$ for sufficiently large $N$.

Next, the legitimate receiver moves to the block $n-1$. Now it already has $\check{a}_{n}$, hence we also have $\check{m}_{n-1}=f(\check{a}_{n})$
(here $f$ is a deterministic function, which means that $\check{m}_{n-1}$ can be determined by $\check{a}_{n}$). It first
declares that $\check{a}_{n-1}$ is received, if $\check{a}_{n-1}$ is the unique one such that
$(x_{r}^{N}(\check{a}_{n-1}),y^{N}(n-1))$ are joint typical. If (\ref{appen6.5}) is satisfied,
$\check{a}_{n-1}=a_{n-1}$ with high probability. After knowing $\check{a}_{n-1}$, the destination gets an estimation of $w_{2,n-1}$
by picking the unique $\check{w}_{2,n-1}$ such that
$(x_{2}^{N}(\check{w}_{2,n-1}),\hat{y}_{r}^{N}(\check{m}_{n-1},\check{a}_{n-1}),y^{N}(n-1),x_{r}^{N}(\check{a}_{n-1}))$ are jointly typical.
We will have $\check{w}_{2,n-1}=w_{2,n-1}$ with high probability, if the codewords of $x_{2}^{N}$ is upper bounded by
$2^{NI(X_{2};Y,\hat{Y}_{r}|X_{r})}$ and $N$ is sufficiently large.

After decoding $\check{w}_{2,n-1}$, the legitimate receiver tries to find a quintuple such that
\\$(x_{1}^{N}(\check{w}_{1,n-1}),
x_{2}^{N}(\check{w}_{2,n-1}),\hat{y}_{r}^{N}(\check{m}_{n-1},\check{a}_{n-1}),y^{N}(n-1),x_{r}^{N}(\check{a}_{n-1}))$ are jointly typical.
Based on the AEP, the probability
$Pr\{\check{w}_{1,n-1}=w_{1,n-1}\}$ goes to $1$ if the codewords of $x_{1}^{N}$ is upper bounded by
$2^{NI(X_{1};Y,\hat{Y}_{r}|X_{2},X_{r})}$ and $N$ is sufficiently large.

The decoding scheme of the legitimate receiver in block $i$ ($1\leq i\leq n-2$) is similar to that in block $n-1$, and we omit it
here.

\item \textbf{Decoding for case 2:}

(At the relay) The relay does not need to decode any codeword.

(At the legitimate receiver) In block $i$ ($1\leq i\leq n$), the legitimate receiver decodes $a_{i}$ by choosing unique
$\check{a}_{i}$ such that $(x_{r}^{N}(\check{a}_{i}),y^{N}(i))$ are jointly typical. Since
$R^{*}_{r2}$ satisfies (\ref{appen6.8}), the probability $Pr\{\check{a}_{i}=a_{i}\}$ goes to $1$ for sufficiently large $N$.

Now since the legitimate receiver has $\check{a}_{i}$, he also knows $\hat{y}_{r}^{N}(\check{a}_{i})$.
Then he gets an estimation of $w_{2,i}$
by picking the unique $\check{w}_{2,i}$ such that
$(x_{2}^{N}(\check{w}_{2,i}),\hat{y}_{r}^{N}(\check{a}_{i}),y^{N}(i),x_{r}^{N}(\check{a}_{i}))$ are jointly typical.
We will have $\check{w}_{2,i}=w_{2,i}$ with high probability, if the codewords of $x_{2}^{N}$ is upper bounded by
$2^{NI(X_{2};Y,\hat{Y}_{r}|X_{r})}$ and $N$ is sufficiently large.

After decoding $\check{w}_{2,i}$, the legitimate receiver tries to find a quintuple such that \\
$(x_{1}^{N}(\check{w}_{1,i}),
x_{2}^{N}(\check{w}_{2,i}),\hat{y}_{r}^{N}(\check{a}_{i}),y^{N}(i),x_{r}^{N}(\check{a}_{i}))$ are jointly typical.
Based on the AEP, the probability
$Pr\{\check{w}_{1,i}=w_{1,i}\}$ goes to $1$ if the codewords of $x_{1}^{N}$ is upper bounded by
$2^{NI(X_{1};Y,\hat{Y}_{r}|X_{2},X_{r})}$ and $N$ is sufficiently large.

\end{itemize}

$P_{e}\leq\epsilon$ is easy to be checked by using the above encoding-decoding schemes.
Now, it remains to prove $\Delta\geq R_{1}+R_{2}-\epsilon$ for both cases, see the followings.

\textbf{Equivocation Analysis:}

\textbf{Proof of $\Delta\geq R_{1}+R_{2}-\epsilon$ for case 1:}

\begin{eqnarray}\label{appen6.12}
\Delta&=&\frac{1}{N}H(W_{1},W_{2}|Z^{N})\nonumber\\
&=&\frac{1}{N}(H(W_{1}|Z^{N})+H(W_{2}|W_{1},Z^{N})).
\end{eqnarray}
The first term in (\ref{appen6.12}) is bounded as follows.

\begin{eqnarray}\label{appen6.13}
\frac{1}{N}H(W_{1}|Z^{N})&=&\frac{1}{N}(H(W_{1},Z^{N})-H(Z^{N}))\nonumber\\
&=&\frac{1}{N}(H(W_{1},Z^{N},X_{1}^{N},X_{r}^{N})-H(X_{1}^{N},X_{r}^{N}|W_{1},Z^{N})-H(Z^{N}))\nonumber\\
&\stackrel{(a)}=&\frac{1}{N}(H(Z^{N}|X_{1}^{N},X_{r}^{N})+H(X_{1}^{N})+H(X_{r}^{N})-H(X_{1}^{N},X_{r}^{N}|W_{1},Z^{N})-H(Z^{N}))\nonumber\\
&=&\frac{1}{N}(H(X_{1}^{N})+H(X_{r}^{N})-I(X_{1}^{N},X_{r}^{N};Z^{N})-H(X_{1}^{N},X_{r}^{N}|W_{1},Z^{N})),
\end{eqnarray}
where (a) follows from $W_{1}\rightarrow (X_{1}^{N},X_{r}^{N})\rightarrow Z^{N}$, $H(W_{1}|X_{1}^{N})=0$ and the fact that $X_{1}^{N}$
is independent of $X_{r}^{N}$.

Consider the first term in (\ref{appen6.13}), the code-book generation of $x_{1}^{N}$ shows that the total number of $x_{1}^{N}$ is
upper bounded by (\ref{appen6.72}). Thus,
using the same approach as that in \cite[Lemma 3]{LP}, we have
\begin{equation}\label{appen6.14}
\frac{1}{N}H(X_{1}^{N})\geq I(X_{1};Y,\hat{Y}_{r}|X_{2},X_{r})+R^{*}-R^{*}_{r1}-\epsilon^{'}-\epsilon_{1,N},
\end{equation}
where $\epsilon_{1,N}\rightarrow 0$ as $N\rightarrow \infty$.

For the second term in (\ref{appen6.13}), the code-book generation of $x_{r}^{N}$ and \cite[Lemma 3]{LP} guarantee that
\begin{equation}\label{appen6.15}
\frac{1}{N}H(X_{r}^{N})\geq R^{*}_{r1}-\epsilon^{'}-\epsilon_{2,N},
\end{equation}
where $\epsilon_{2,N}\rightarrow 0$ as $N\rightarrow \infty$.

For the third term in (\ref{appen6.13}), since the channel is memoryless, and $X_{1}^{N}$, $X_{2}^{N}$, $X_{r}^{N}$ are i.i.d. generated, we get
\begin{equation}\label{appen6.16}
\frac{1}{N}I(X_{1}^{N},X_{r}^{N};Z^{N})=I(X_{1},X_{r};Z).
\end{equation}

Now, we consider the last term of (\ref{appen6.13}). Given $w_{1}$, the
wiretapper can do joint decoding. Specifically, given
$z^{N}$ and $w_{1}$,
\begin{equation}\label{appen6.17}
\frac{1}{N}H(X_{1}^{N},X_{r}^{N}|W_{1},Z^{N})\leq \epsilon_{3,N}
\end{equation}
($\epsilon_{3,N}\rightarrow 0$ as $N\rightarrow \infty$) is guaranteed if
$R_{r}\leq I(X_{r};Z|X_{1})$ and $I(X_{1};Y,\hat{Y}_{r}|X_{2},X_{r})-\epsilon^{'}+R^{*}-R^{*}_{r1}-R_{1}\leq I(X_{1};Z|X_{r})$, and
this is from the properties of AEP (similar argument is used
in the proof of \cite[Theorem 3]{LG}). By checking (\ref{appen6.5}) and (\ref{appen6.72}),
(\ref{appen6.17}) is obtained.

Substituting (\ref{appen6.14}), (\ref{appen6.15}), (\ref{appen6.16}) and (\ref{appen6.17}) into (\ref{appen6.13}), we have
\begin{equation}\label{appen6.18}
\frac{1}{N}H(W_{1}|Z^{N})\geq I(X_{1};Y,\hat{Y}_{r}|X_{2},X_{r})+R^{*}-I(X_{1},X_{r};Z)-2\epsilon^{'}-\epsilon_{1,N}-\epsilon_{2,N}-\epsilon_{3,N}.
\end{equation}

The second term in (\ref{appen6.12}) is bounded as follows.

\begin{eqnarray}\label{appen6.19}
\frac{1}{N}H(W_{2}|W_{1},Z^{N})&\geq&\frac{1}{N}H(W_{2}|W_{1},Z^{N},X_{1}^{N},X_{r}^{N})\nonumber\\
&\stackrel{(1)}=&\frac{1}{N}H(W_{2}|Z^{N},X_{1}^{N},X_{r}^{N})\nonumber\\
&=&\frac{1}{N}(H(W_{2},Z^{N},X_{1}^{N},X_{r}^{N})-H(Z^{N},X_{1}^{N},X_{r}^{N}))\nonumber\\
&=&\frac{1}{N}(H(W_{2},Z^{N},X_{1}^{N},X_{r}^{N},X_{2}^{N})-H(X_{2}^{N}|W_{2},Z^{N},X_{1}^{N},X_{r}^{N})-H(Z^{N},X_{1}^{N},X_{r}^{N}))\nonumber\\
&\stackrel{(2)}=&\frac{1}{N}(H(Z^{N}|X_{1}^{N},X_{2}^{N},X_{r}^{N})+H(X_{r}^{N})+H(X_{1}^{N})+H(X_{2}^{N})\nonumber\\
&&-H(X_{2}^{N}|W_{2},Z^{N},X_{1}^{N},X_{r}^{N})-H(Z^{N}|X_{1}^{N},X_{r}^{N})-H(X_{1}^{N})-H(X_{r}^{N}))\nonumber\\
&=&\frac{1}{N}(H(X_{2}^{N})-I(X_{2}^{N};Z^{N}|X_{1}^{N},X_{r}^{N})-H(X_{2}^{N}|W_{2},Z^{N},X_{1}^{N},X_{r}^{N})),
\end{eqnarray}
where (1) is from the Markov chain $W_{1}\rightarrow (Z^{N},X_{1}^{N},X_{r}^{N})\rightarrow W_{2}$, and (2) is from
the Markov chain $W_{2}\rightarrow (X_{1}^{N},X_{2}^{N},X_{r}^{N})\rightarrow Z^{N}$, $H(W_{2}|X_{2}^{N})=0$, and the fact that
$X_{1}^{N}$, $X_{2}^{N}$ and $X_{r}^{N}$ are independent.

Consider the first term in (\ref{appen6.19}),
using the same approach as that in \cite[Lemma 3]{LP}, we have
\begin{equation}\label{appen6.20}
\frac{1}{N}H(X_{2}^{N})\geq I(X_{2};Y,\hat{Y}_{r}|X_{r})-\epsilon^{'}-\epsilon_{4,N},
\end{equation}
where $\epsilon_{4,N}\rightarrow 0$ as $N\rightarrow \infty$.

For the second term in (\ref{appen6.19}), since the channel is memoryless, and $X_{1}^{N}$, $X_{2}^{N}$, $X_{r}^{N}$ are i.i.d. generated, we get
\begin{equation}\label{appen6.21}
\frac{1}{N}I(X_{2}^{N};Z^{N}|X_{1}^{N},X_{r}^{N})=I(X_{2};Z|X_{1},X_{r}).
\end{equation}

Now, we consider the last term of (\ref{appen6.19}). Given $Z^{N}$, $X_{1}^{N}$, $X_{r}^{N}$ and $W_{2}$,
the total number of possible codewords of $x_{2}^{N}$ is $2^{N(I(X_{2};Y,\hat{Y}_{r}|X_{r})-\epsilon^{'}-R_{2})}$.
By using the Fano's inequality and (\ref{appen6.7}),
we have
\begin{equation}\label{appen6.22}
\frac{1}{N}H(X_{2}^{N}|W_{2},Z^{N},X_{1}^{N},X_{r}^{N})\leq \epsilon_{5,N}.
\end{equation}

Substituting (\ref{appen6.20}), (\ref{appen6.21}) and (\ref{appen6.22}) into (\ref{appen6.19}), we have
\begin{equation}\label{appen6.23}
\frac{1}{N}H(W_{2}|W_{1},Z^{N})\geq I(X_{2};Y,\hat{Y}_{r}|X_{r})-I(X_{2};Z|X_{1},X_{r})-\epsilon^{'}-\epsilon_{4,N}-\epsilon_{5,N}.
\end{equation}

Substituting (\ref{appen6.18}) and (\ref{appen6.23}) into (\ref{appen6.12}),
and choosing $\epsilon^{'}$ and sufficiently large $N$ such that $3\epsilon^{'}+\epsilon_{1,N}+\epsilon_{2,N}+\epsilon_{3,N}+\epsilon_{4,N}+\epsilon_{5,N}\leq \epsilon$,
$\Delta\geq R_{1}+R_{2}-\epsilon$ for case 1 is proved.

\textbf{Proof of $\Delta\geq R_{1}+R_{2}-\epsilon$ for case 2:}

\begin{eqnarray}\label{appen6.24}
\Delta&=&\frac{1}{N}H(W_{1},W_{2}|Z^{N})\nonumber\\
&=&\frac{1}{N}(H(W_{1}|Z^{N})+H(W_{2}|W_{1},Z^{N})).
\end{eqnarray}
The first term in (\ref{appen6.24}) is bounded as follows.

\begin{eqnarray}\label{appen6.25}
\frac{1}{N}H(W_{1}|Z^{N})&\geq&\frac{1}{N}H(W_{1}|Z^{N},X_{r}^{N})\nonumber\\
&=&\frac{1}{N}(H(W_{1},Z^{N},X_{r}^{N})-H(Z^{N},X_{r}^{N}))\nonumber\\
&=&\frac{1}{N}(H(W_{1},Z^{N},X_{1}^{N},X_{r}^{N})-H(X_{1}^{N}|W_{1},Z^{N},X_{r}^{N})-H(Z^{N},X_{r}^{N}))\nonumber\\
&\stackrel{(a)}=&\frac{1}{N}(H(Z^{N}|X_{1}^{N},X_{r}^{N})+H(X_{1}^{N})+H(X_{r}^{N})-H(X_{1}^{N}|W_{1},Z^{N},X_{r}^{N})\nonumber\\
&&-H(Z^{N}|X_{r}^{N})-H(X_{r}^{N}))\nonumber\\
&=&\frac{1}{N}(H(X_{1}^{N})-I(X_{1}^{N};Z^{N}|X_{r}^{N})-H(X_{1}^{N}|W_{1},Z^{N},X_{r}^{N})),
\end{eqnarray}
where (a) follows from $W_{1}\rightarrow (X_{1}^{N},X_{r}^{N})\rightarrow Z^{N}$, $H(W_{1}|X_{1}^{N})=0$ and the fact that $X_{1}^{N}$
is independent of $X_{r}^{N}$.

Consider the first term in (\ref{appen6.25}), the code-book generation of $x_{1}^{N}$ shows that the total number of $x_{1}^{N}$ is
$2^{N(I(X_{1};Y,\hat{Y}_{r}|X_{2},X_{r})-\epsilon^{'})}$. Thus,
using the same approach as that in \cite[Lemma 3]{LP}, we have
\begin{equation}\label{appen6.26}
\frac{1}{N}H(X_{1}^{N})\geq I(X_{1};Y,\hat{Y}_{r}|X_{2},X_{r})-\epsilon^{'}-\epsilon_{1,N},
\end{equation}
where $\epsilon_{1,N}\rightarrow 0$ as $N\rightarrow \infty$.

For the second term in (\ref{appen6.25}), since the channel is memoryless, and $X_{1}^{N}$, $X_{2}^{N}$, $X_{r}^{N}$ are i.i.d. generated, we get
\begin{equation}\label{appen6.27}
\frac{1}{N}I(X_{1}^{N};Z^{N}|X_{r}^{N})=I(X_{1};Z|X_{r}).
\end{equation}

Now, we consider the last term of (\ref{appen6.25}). Given $Z^{N}$, $X_{r}^{N}$ and $W_{1}$,
the total number of possible codewords of $x_{1}^{N}$ is $2^{N(I(X_{1};Y,\hat{Y}_{r}|X_{2},X_{r})-\epsilon^{'}-R_{1})}$.
By using the Fano's inequality and (\ref{appen6.10}),
we have
\begin{equation}\label{appen6.28}
\frac{1}{N}H(X_{1}^{N}|W_{1},Z^{N},X_{r}^{N})\leq \epsilon_{2,N},
\end{equation}
where $\epsilon_{2,N}\rightarrow 0$ as $N\rightarrow \infty$.

Substituting (\ref{appen6.26}), (\ref{appen6.27}) and (\ref{appen6.28}) into (\ref{appen6.25}), we have
\begin{equation}\label{appen6.29}
\frac{1}{N}H(W_{1}|Z^{N})\geq I(X_{1};Y,\hat{Y}_{r}|X_{2},X_{r})-I(X_{1};Z|X_{r})-\epsilon^{'}-\epsilon_{1,N}-\epsilon_{2,N}.
\end{equation}

The second term in (\ref{appen6.24}) is bounded the same as that for case 1, and thus, we have
\begin{equation}\label{appen6.30}
\lim_{N\rightarrow \infty}\frac{1}{N}H(W_{2}|W_{1},Z^{N})\geq I(X_{2};Y,\hat{Y}_{r}|X_{r})-I(X_{2};Z|X_{1},X_{r})-\epsilon^{'}-\epsilon_{3,N}-\epsilon_{4,N}.
\end{equation}
The proof is omitted here.

Substituting (\ref{appen6.29}) and (\ref{appen6.30}) into (\ref{appen6.24}),
and choosing $\epsilon^{'}$ and sufficiently large $N$ such that $2\epsilon^{'}+\epsilon_{1,N}+\epsilon_{2,N}+\epsilon_{3,N}+\epsilon_{4,N}\leq \epsilon$,
$\Delta\geq R_{1}+R_{2}-\epsilon$ for case 2 is proved.

The proof of Theorem \ref{T3} is completed.

\section{Proof of Theorem \ref{T4.1}\label{appen4}}

In this section, we prove Theorem \ref{T4.1}: all the achievable secrecy pairs
$(R_{1},R_{2})$ of the degraded discrete memoryless MARC-WT are contained in the set $\mathcal{R}^{ddo}$.
We will prove the inequalities of Theorem \ref{T4.1} in the remainder of this section.

\textbf{(Proof of $R_{1}\leq I(X_{1},X_{r};Y|X_{2},U)-I(X_{1};Z|U)$)}:

\begin{eqnarray}\label{b1}
R_{1}-\epsilon&=&\frac{1}{N}H(W_{1})-\epsilon\stackrel{(1)}\leq\frac{1}{N}H(W_{1}|Z^{N})\nonumber\\
&\stackrel{(2)}\leq&\frac{1}{N}(H(W_{1}|Z^{N})-H(W_{1}|Z^{N},W_{2},Y^{N},X_{2}^{N})+\delta(P_{e}))\nonumber\\
&\stackrel{(3)}=&\frac{1}{N}(H(W_{1}|Z^{N})-H(W_{1}|Z^{N},Y^{N},X_{2}^{N})+\delta(P_{e}))\nonumber\\
&=&\frac{1}{N}(I(W_{1};Y^{N},X_{2}^{N}|Z^{N})+\delta(P_{e}))\nonumber\\
&\leq&\frac{1}{N}(H(Y^{N},X_{2}^{N}|Z^{N})-H(Y^{N},X_{2}^{N}|Z^{N},W_{1},X_{1}^{N})+\delta(P_{e}))\nonumber\\
&\stackrel{(4)}=&\frac{1}{N}(H(Y^{N},X_{2}^{N}|Z^{N})-H(Y^{N},X_{2}^{N}|Z^{N},X_{1}^{N})+\delta(P_{e}))\nonumber\\
&=&\frac{1}{N}(I(Y^{N},X_{2}^{N};X_{1}^{N}|Z^{N})+\delta(P_{e}))\nonumber\\
&\stackrel{(5)}=&\frac{1}{N}(H(X_{1}^{N}|Z^{N})-H(X_{1}^{N}|Z^{N},Y^{N},X_{2}^{N})-H(X_{1}^{N})+H(X_{1}^{N}|X_{2}^{N})+\delta(P_{e}))\nonumber\\
&\stackrel{(6)}=&\frac{1}{N}(I(X_{1}^{N};Y^{N}|X_{2}^{N})-I(X_{1}^{N};Z^{N})+\delta(P_{e}))\nonumber\\
&\leq&\frac{1}{N}(I(X_{1}^{N},X_{r}^{N};Y^{N}|X_{2}^{N})-I(X_{1}^{N};Z^{N})+\delta(P_{e}))\nonumber\\
&=&\frac{1}{N}\sum_{i=1}^{N}(H(Y_{i}|Y^{i-1},X_{2}^{N})-H(Y_{i}|X_{1,i},X_{2,i},X_{r,i})-H(Z_{i}|Z^{i-1})+H(Z_{i}|Z^{i-1},X_{1}^{N}))+\frac{\delta(P_{e})}{N}\nonumber\\
&\stackrel{(7)}=&\frac{1}{N}\sum_{i=1}^{N}(H(Y_{i}|Y^{i-1},X_{2}^{N},Z^{i-1})-H(Y_{i}|X_{1,i},X_{2,i},X_{r,i},Z^{i-1})-H(Z_{i}|Z^{i-1})+H(Z_{i}|Z^{i-1},X_{1}^{N}))+\frac{\delta(P_{e})}{N}\nonumber\\
&\leq&\frac{1}{N}\sum_{i=1}^{N}(H(Y_{i}|X_{2,i},Z^{i-1})-H(Y_{i}|X_{1,i},X_{2,i},X_{r,i},Z^{i-1})-H(Z_{i}|Z^{i-1})+H(Z_{i}|Z^{i-1},X_{1,i}))+\frac{\delta(P_{e})}{N}\nonumber\\
&\stackrel{(8)}=&\frac{1}{N}\sum_{i=1}^{N}(H(Y_{i}|X_{2,i},Z^{i-1},J=i)-H(Y_{i}|X_{1,i},X_{2,i},X_{r,i},Z^{i-1},J=i)-H(Z_{i}|Z^{i-1},J=i)\nonumber\\
&&+H(Z_{i}|Z^{i-1},X_{1,i},J=i))+\frac{\delta(P_{e})}{N}\nonumber\\
&\stackrel{(9)}=&H(Y_{J}|X_{2,J},Z^{J-1},J)-H(Y_{J}|X_{1,J},X_{2,J},X_{r,J},Z^{J-1},J)-H(Z_{J}|Z^{J-1},J)+H(Z_{J}|Z^{J-1},X_{1,J},J)+\frac{\delta(P_{e})}{N}\nonumber\\
&\stackrel{(10)}=&I(X_{1},X_{r};Y|X_{2},U)-I(X_{1};Z|U)+\frac{\delta(P_{e})}{N},
\end{eqnarray}
where (1) is from the fact that the secrecy requirement on the full message set also ensures the secrecy of
individual message (see (\ref{e202.x})), (2) is from the Fano¡¯s inequality, (3) is from $H(W_{2}|X_{2}^{N})=0$,
(4) is from $H(W_{1}|X_{1}^{N})=0$, (5) and (6) are from the fact that the wiretap channel is degraded, which implies the
Markov chain $X_{1}^{N}\rightarrow (X_{2}^{N},Y^{N})\rightarrow Z^{N}$,
and from the fact that $X_{1}^{N}$ is independent of $X_{2}^{N}$,
(7) is from the Markov chains
$Y_{i}\rightarrow (Y^{i-1},X_{2}^{N})\rightarrow Z^{i-1}$ and $Y_{i}\rightarrow (X_{1,i},X_{2,i},X_{r,i})\rightarrow Z^{i-1}$
(these Markov chains are also from the fact that the wiretap channel is degraded),
(8) is from $J$ is a random variable (uniformly distributed over $\{1,2,...,N\}$), and it is independent of
$X_{1}^{N}$, $X_{2}^{N}$, $X_{r}^{N}$, $Y^{N}$ and $Z^{N}$, (9) is from $J$ is uniformly distributed over $\{1,2,...,N\}$,
and (10) is from the definitions that $X_{1}\triangleq X_{1,J}$, $X_{2}\triangleq X_{2,J}$, $X_{r}\triangleq X_{r,J}$, $Y\triangleq Y_{J}$,
$Z\triangleq Z_{J}$ and $U\triangleq (Z^{J-1},J)$.

By using $P_{e}\leq \epsilon$ and letting $\epsilon\rightarrow 0$, $R_{1}\leq I(X_{1},X_{r};Y|X_{2},U)-I(X_{1};Z|U)$
is proved.

\textbf{(Proof of $R_{2}\leq I(X_{2},X_{r};Y|X_{1},U)-I(X_{2};Z|U)$)}:

The proof is analogous to the proof of $R_{1}\leq I(X_{1},X_{r};Y|X_{2},U)-I(X_{1};Z|U)$, and it is omitted here.

\textbf{Proof of $R_{1}+R_{2}\leq I(X_{1},X_{2},X_{r};Y|U)-I(X_{1},X_{2};Z|U)$}:

\begin{eqnarray}\label{b2}
R_{1}+R_{2}-\epsilon&\stackrel{(1)}\leq&\Delta=\frac{1}{N}H(W_{1},W_{2}|Z^{N})\nonumber\\
&\stackrel{(2)}\leq&\frac{1}{N}(H(W_{1},W_{2}|Z^{N})+\delta(P_{e})-H(W_{1},W_{2}|Y^{N},Z^{N}))\nonumber\\
&\leq&\frac{1}{N}(H(Y^{N}|Z^{N})-H(Y^{N}|Z^{N},W_{1},W_{2},X_{1}^{N},X_{2}^{N})+\delta(P_{e}))\nonumber\\
&\stackrel{(3)}=&\frac{1}{N}(H(Y^{N}|Z^{N})-H(Y^{N}|Z^{N},X_{1}^{N},X_{2}^{N})+\delta(P_{e}))\nonumber\\
&=&\frac{1}{N}(I(X_{1}^{N},X_{2}^{N};Y^{N})-I(X_{1}^{N},X_{2}^{N};Z^{N})+\delta(P_{e}))\nonumber\\
&\leq&\frac{1}{N}(I(X_{1}^{N},X_{2}^{N},X_{r}^{N};Y^{N})-I(X_{1}^{N},X_{2}^{N};Z^{N})+\delta(P_{e}))\nonumber\\
&\stackrel{(4)}=&\frac{1}{N}\sum_{i=1}^{N}(H(Y_{i}|Y^{i-1})-H(Y_{i}|X_{1,i},X_{2,i},X_{r,i},Z^{i-1})
-H(Z_{i}|Z^{i-1})+H(Z_{i}|X_{1,i},X_{2,i},Z^{i-1}))+\frac{\delta(P_{e})}{N}\nonumber\\
&\stackrel{(5)}\leq&\frac{1}{N}\sum_{i=1}^{N}(H(Y_{i}|Z^{i-1})-H(Y_{i}|X_{1,i},X_{2,i},X_{r,i},Z^{i-1})
-H(Z_{i}|Z^{i-1})+H(Z_{i}|X_{1,i},X_{2,i},Z^{i-1}))+\frac{\delta(P_{e})}{N}\nonumber\\
&\stackrel{(6)}=&\frac{1}{N}\sum_{i=1}^{N}(H(Y_{i}|Z^{i-1},J=i)-H(Y_{i}|X_{1,i},X_{2,i},X_{r,i},Z^{i-1},J=i)\nonumber\\
&&-H(Z_{i}|Z^{i-1},J=i)+H(Z_{i}|X_{1,i},X_{2,i},Z^{i-1},J=i))+\frac{\delta(P_{e})}{N}\nonumber\\
&\stackrel{(7)}=&H(Y_{J}|Z^{J-1},J)-H(Y_{J}|X_{1,J},X_{2,J},X_{r,J},Z^{J-1},J)\nonumber\\
&&-H(Z_{J}|Z^{J-1},J)+H(Z_{J}|X_{1,J},X_{2,J},Z^{J-1},J)+\frac{\delta(P_{e})}{N}\nonumber\\
&\stackrel{(8)}\leq &I(X_{1},X_{2},X_{r};Y|U)-I(X_{1},X_{2};Z|U)+\frac{\delta(\epsilon)}{N},
\end{eqnarray}
where (1) is from (\ref{e202}),
(2) is from the Fano¡¯s inequality, (3) is from $(W_{1},W_{2})\rightarrow (X_{1}^{N},X_{2}^{N},Z^{N})\rightarrow Y^{N}$,
(4) is from $Y_{i}\rightarrow (X_{1,i},X_{2,i},X_{r,i})\rightarrow Z^{i-1}$,
(5) is from $Y_{i}\rightarrow Y^{i-1}\rightarrow Z^{i-1}$, (6) is from $J$ is a
random variable (uniformly distributed over $\{1,2,...,N\}$), and it is independent of
$X_{1}^{N}$, $X_{2}^{N}$, $X_{r}^{N}$, $Y^{N}$ and $Z^{N}$, (7) is from $J$ is uniformly distributed over $\{1,2,...,N\}$,
and (8) is from the definitions that $X_{1}\triangleq X_{1,J}$, $X_{2}\triangleq X_{2,J}$, $X_{r}\triangleq X_{r,J}$, $Y\triangleq Y_{J}$,
$Z\triangleq Z_{J}$ and $U\triangleq (Z^{J-1},J)$, and the fact that
$P_{e}\leq \epsilon$.

Letting $\epsilon\rightarrow 0$, $R_{1}+R_{2}\leq I(X_{1},X_{2},X_{r};Y|U)-I(X_{1},X_{2};Z|U)$ is proved.

The proof of Theorem \ref{T4.1} is completed.

\section{Proof of Theorem \ref{T6.5}\label{appen5}}

Since $N_{2}\geq N_{1}$, the GMARC-WT reduces to a kind of degraded MARC-WT with the Markov chain $(X_{1},X_{2},X_{r},Y_{r})\rightarrow Y\rightarrow Z$,
and thus the outer bound $\mathcal{R}^{gout}$
can be obtained from Theorem \ref{T4.1}. The details are as follows.

From (\ref{b1}), we know that
\begin{eqnarray}\label{bh1}
R_{1}&\leq&\frac{1}{N}\sum_{i=1}^{N}(h(Y_{i}|X_{2,i},Z^{i-1})-h(Y_{i}|X_{1,i},X_{2,i},X_{r,i},Z^{i-1})\nonumber\\
&&-h(Z_{i}|Z^{i-1})+h(Z_{i}|Z^{i-1},X_{1,i}))+\frac{\delta(P_{e})}{N}.
\end{eqnarray}

Analogously,
\begin{eqnarray}\label{bh2}
R_{2}&\leq&\frac{1}{N}\sum_{i=1}^{N}(h(Y_{i}|X_{1,i},Z^{i-1})-h(Y_{i}|X_{1,i},X_{2,i},X_{r,i},Z^{i-1})-h(Z_{i}|Z^{i-1})\nonumber\\
&&+h(Z_{i}|Z^{i-1},X_{2,i}))+\frac{\delta(P_{e})}{N}.
\end{eqnarray}

From (\ref{b2}), we have
\begin{eqnarray}\label{bh3}
R_{1}+R_{2}&\leq&\frac{1}{N}\sum_{i=1}^{N}(h(Y_{i}|Z^{i-1})-h(Y_{i}|X_{1,i},X_{2,i},X_{r,i},Z^{i-1})-h(Z_{i}|Z^{i-1})\nonumber\\
&&+h(Z_{i}|X_{1,i},X_{2,i},Z^{i-1}))+\frac{\delta(P_{e})}{N}).
\end{eqnarray}

It remains to bound the conditional entropies in (\ref{bh1}), (\ref{bh2}) and (\ref{bh3}), see the followings.

First note that
\begin{eqnarray}\label{bh4}
\frac{1}{N}\sum_{i=1}^{N}h(Z_{i}|X_{1,i},X_{2,i},Z^{i-1})&\leq&\frac{1}{N}\sum_{i=1}^{N}h(Z_{i}|X_{1,i},X_{2,i})\nonumber\\
&\stackrel{(1)}\leq&\frac{1}{N}\sum_{i=1}^{N}h(Z_{2,i}+X_{r,i})\nonumber\\
&\leq& \frac{1}{N}\sum_{i=1}^{N}\frac{1}{2}\log2\pi e(E[X_{r,i}^{2}]+N_{2})\nonumber\\
&\stackrel{(2)}\leq& \frac{1}{2}\log2\pi e(\frac{1}{N}\sum_{i=1}^{N}E[X_{r,i}^{2}]+N_{2})\nonumber\\
&\leq& \frac{1}{2}\log2\pi e(P_{r}+N_{2}),
\end{eqnarray}
where (1) is from $Z_{i}=X_{1,i}+X_{2,i}+X_{r,i}+Z_{2,i}$, and (2) is from Jensen's inequality.

On the other hand,
\begin{eqnarray}\label{bh5}
\frac{1}{N}\sum_{i=1}^{N}h(Z_{i}|X_{1,i},X_{2,i},Z^{i-1})&\geq&\frac{1}{N}\sum_{i=1}^{N}h(Z_{i}|X_{1,i},X_{2,i},X_{r,i},Z^{i-1})\nonumber\\
&\stackrel{(a)}=&\frac{1}{N}\sum_{i=1}^{N}h(Z_{i}|X_{1,i},X_{2,i},X_{r,i})\nonumber\\
&=&\frac{1}{N}\sum_{i=1}^{N}h(Z_{2,i})\nonumber\\
&=&\frac{1}{N}\sum_{i=1}^{N}\frac{1}{2}\log2\pi e N_{2}=\frac{1}{2}\log2\pi e N_{2},
\end{eqnarray}
where (a) is from the Markov chain $Z^{i-1}\rightarrow (X_{1,i},X_{2,i},X_{r,i})\rightarrow Z_{i}$.

Combining (\ref{bh4}) and (\ref{bh5}), we establish that there exists
some $\alpha\in [0,1]$ such that
\begin{eqnarray}\label{bh6}
&&\frac{1}{N}\sum_{i=1}^{N}h(Z_{i}|X_{1,i},X_{2,i},Z^{i-1})=\frac{1}{2}\log2\pi e(\alpha P_{r}+N_{2}).
\end{eqnarray}

Second, since
\begin{eqnarray}\label{bh7}
\frac{1}{N}\sum_{i=1}^{N}h(Z_{i}|X_{1,i},Z^{i-1})&\geq&\frac{1}{N}\sum_{i=1}^{N}h(Z_{i}|X_{1,i},X_{2,i},Z^{i-1})\nonumber\\
&=&\frac{1}{2}\log2\pi e(\alpha P_{r}+N_{2}),
\end{eqnarray}
and
\begin{eqnarray}\label{bh8}
\frac{1}{N}\sum_{i=1}^{N}h(Z_{i}|X_{1,i},Z^{i-1})&\leq&\frac{1}{N}\sum_{i=1}^{N}h(Z_{i}|X_{1,i})\nonumber\\
&\leq&\frac{1}{N}\sum_{i=1}^{N}h(Z_{2,i}+X_{2,i}+X_{r,i})\nonumber\\
&\leq&\frac{1}{N}\sum_{i=1}^{N}\frac{1}{2}\log2\pi e(E[X_{r,i}^{2}]+E[X_{2,i}^{2}]+N_{2})\nonumber\\
&\leq&\frac{1}{2}\log2\pi e(\frac{1}{N}\sum_{i=1}^{N}E[X_{r,i}^{2}]+\frac{1}{N}\sum_{i=1}^{N}E[X_{2,i}^{2}]+N_{2})\nonumber\\
&\leq&\frac{1}{2}\log2\pi e(P_{r}+P_{2}+N_{2}),
\end{eqnarray}
we establish that there exists some $\beta_{1}\in [0,1]$ such that
\begin{eqnarray}\label{bh9}
&&\frac{1}{N}\sum_{i=1}^{N}h(Z_{i}|X_{1,i},Z^{i-1})=\frac{1}{2}\log2\pi e(\alpha P_{r}+N_{2}+\beta_{1}(P_{r}+P_{2}+N_{2}-\alpha P_{r}-N_{2}))\nonumber\\
&&=\frac{1}{2}\log2\pi e(N_{2}+P_{r}(\alpha+\beta_{1}-\alpha \beta_{1})+\beta_{1}P_{2}).
\end{eqnarray}

Third, analogously, there exists some $\beta_{2}\in [0,1]$ such that
\begin{eqnarray}\label{bh10}
&&\frac{1}{N}\sum_{i=1}^{N}h(Z_{i}|X_{2,i},Z^{i-1})=\frac{1}{2}\log2\pi e(N_{2}+P_{r}(\alpha+\beta_{2}-\alpha \beta_{2})+\beta_{2}P_{1}).
\end{eqnarray}

Fourth, since
\begin{eqnarray}\label{bh11}
\frac{1}{N}\sum_{i=1}^{N}h(Z_{i}|Z^{i-1})&\geq&\frac{1}{N}\sum_{i=1}^{N}h(Z_{i}|X_{1,i},Z^{i-1})\nonumber\\
&=&\frac{1}{2}\log2\pi e(N_{2}+P_{r}(\alpha+\beta_{1}-\alpha \beta_{1})+\beta_{1}P_{2}),
\end{eqnarray}

\begin{eqnarray}\label{bh12}
\frac{1}{N}\sum_{i=1}^{N}h(Z_{i}|Z^{i-1})&\geq&\frac{1}{N}\sum_{i=1}^{N}h(Z_{i}|X_{2,i},Z^{i-1})\nonumber\\
&=&\frac{1}{2}\log2\pi e(N_{2}+P_{r}(\alpha+\beta_{2}-\alpha \beta_{2})+\beta_{2}P_{1})£¬
\end{eqnarray}
and
\begin{eqnarray}\label{bh13}
\frac{1}{N}\sum_{i=1}^{N}h(Z_{i}|Z^{i-1})&\leq&\frac{1}{N}\sum_{i=1}^{N}h(Z_{i})\nonumber\\
&=&\frac{1}{N}\sum_{i=1}^{N}h(Z_{2,i}+X_{1,i}+X_{2,i}+X_{r,i})\nonumber\\
&\leq&\frac{1}{N}\sum_{i=1}^{N}\frac{1}{2}\log2\pi e(E[X_{r,i}^{2}]+E[X_{1,i}^{2}]+E[X_{2,i}^{2}]+N_{2})\nonumber\\
&\leq&\frac{1}{2}\log2\pi e(\frac{1}{N}\sum_{i=1}^{N}E[X_{r,i}^{2}]+\frac{1}{N}\sum_{i=1}^{N}E[X_{1,i}^{2}]+\frac{1}{N}\sum_{i=1}^{N}E[X_{2,i}^{2}]+N_{2})\nonumber\\
&\leq&\frac{1}{2}\log2\pi e(P_{r}+P_{1}+P_{2}+N_{2}),
\end{eqnarray}
there exists some $\gamma\in [0,1]$ such that
\begin{eqnarray}\label{bh14}
&&\frac{1}{N}\sum_{i=1}^{N}h(Z_{i}|Z^{i-1})=\frac{1}{2}\log2\pi e(C+\gamma(P_{r}+P_{1}+P_{2}+N_{2}-C)),
\end{eqnarray}
where $C$ is given by
\begin{eqnarray}\label{bh15}
C&=&\max\{N_{2}+P_{r}(\alpha+\beta_{1}-\alpha\beta_{1})+\beta_{1}P_{2}, N_{2}+P_{r}(\alpha+\beta_{2}-\alpha\beta_{2})+\beta_{2}P_{1}\}.
\end{eqnarray}

Fifth, by using the entropy power inequality, we have
\begin{eqnarray}\label{bh16}
2^{2h(Z_{i}|X_{1,i},Z^{i-1})}&\stackrel{(1)}=&2^{2h(Y_{i}+Z^{'}_{2,i}|X_{1,i},Z^{i-1})}\nonumber\\
&\stackrel{(2)}\geq& 2^{2h(Y_{i}|X_{1,i},Z^{i-1})}+2^{2h(Z^{'}_{2,i}|X_{1,i},Z^{i-1})}\nonumber\\
&\stackrel{(3)}=&2^{2h(Y_{i}|X_{1,i},Z^{i-1})}+2^{2h(Z^{'}_{2,i})},
\end{eqnarray}
where (1) is from the definition that $Z^{'}_{2,i}=Z_{2,i}-Z_{1,i}$, (2) is from the entropy power inequality,
and (3) is from $Z^{'}_{2,i}$ is independent of $X_{1,i}$ and $Z^{i-1}$.

Substituting $h(Z^{'}_{2,i})=\frac{1}{2}\log2\pi e(N_{2}-N_{1})$ and (\ref{bh9}) into (\ref{bh16}), and using Jensen's inequality, we have
\begin{eqnarray}\label{bh17}
&&\frac{1}{N}\sum_{i=1}^{N}h(Y_{i}|X_{1,i},Z^{i-1})\leq \frac{1}{2}\log2\pi e(P_{r}(\alpha+\beta_{1}-\alpha \beta_{1})+\beta_{1}P_{2}+N_{1}).
\end{eqnarray}

Analogously, we have
\begin{eqnarray}\label{bh18}
&&\frac{1}{N}\sum_{i=1}^{N}h(Y_{i}|X_{2,i},Z^{i-1})\leq \frac{1}{2}\log2\pi e(P_{r}(\alpha+\beta_{2}-\alpha \beta_{2})+\beta_{2}P_{1}+N_{1}),
\end{eqnarray}
and
\begin{eqnarray}\label{bh19}
&&\frac{1}{N}\sum_{i=1}^{N}h(Y_{i}|Z^{i-1})\leq \frac{1}{2}\log2\pi e(C+\gamma(P_{r}+P_{1}+P_{2}+N_{1}-C)).
\end{eqnarray}

Finally, note that
\begin{eqnarray}\label{bh20}
h(Y_{i}|X_{1,i},X_{2,i},X_{r,i},Z^{i-1})&=&h(Z_{1,i}|X_{1,i},X_{2,i},X_{r,i},Z^{i-1})\nonumber\\
&\stackrel{(1)}=&h(Z_{1,i})=\frac{1}{2}\log2\pi eN_{1},
\end{eqnarray}
where (1) is from $Z_{1,i}$ is independent of $X_{1,i}$, $X_{2,i}$, $X_{r,i}$ and $Z^{i-1}$.

Substituting (\ref{bh6}), (\ref{bh9}), (\ref{bh10}), (\ref{bh14}), (\ref{bh17}), (\ref{bh18}), (\ref{bh19}) and (\ref{bh20})
into (\ref{bh1}), (\ref{bh2}) and (\ref{bh3}), using the fact that $P_{e}\leq \epsilon$ and letting $\epsilon\rightarrow 0$,
Theorem \ref{T6.5} is proved.

\end{document}